\documentclass{optica-article}

\journal{opticajournal} 

\articletype{Research Article}
\usepackage{stfloats}
\usepackage{lineno}
\usepackage{pdfpages}
\usepackage{comment}

\begin{document}

\title{Adaptive Information Mining: Approaching the Quantum Limit of Super-Resolution Imaging}
\author{Seungman Choi,\authormark{1,†} Peter Menart,\authormark{1,†} Andrew Schramka,\authormark{1} and Zubin Jacob\authormark{1,*}}

\address{\authormark{1}Elmore Family School of Electrical and Computer Engineering, Birck Nanotechnology Center, Purdue University, West Lafayette, Indiana 47907, USA.\\
\authormark{†}
These two authors contributed equally to this work.\\
\email{\authormark{*}zjacob@purdue.edu} 
}

\begin{abstract*}
The "Rayleigh limit" is not an intrinsic property of light, but a consequence of sub-optimal measurement of collected photons by the imaging instrument. To overcome this limitation, we present Adaptive Information Mining, a framework that achieves super-resolution by using sequentially refined measurements for incoming photons.
Our approach employs a reconfigurable imaging system that dynamically reshapes the instrument measurement responses to maximize the Fisher information and guide scene estimation toward the fundamental quantum limit, even in deep sub-Rayleigh regimes. We demonstrate that this approach achieves an estimation error 2-5$\times$ closer to the fundamental quantum limit than traditional direct imaging when imaging point sources, without requiring any prior knowledge of their positions and brightnesses. Importantly, we demonstrate that AIM can be extended to fully blind multi-parameter estimation in scenarios involving an unknown number of sources or turbulence-induced wavefront distortions. Our approach addresses the limitations of conventional fixed imaging, offering a practical path toward quantum-limited super-resolution imaging.


\end{abstract*}


\section{Introduction}
The pursuit of high-resolution astronomical imaging---critical for exoplanet detection \cite{sauvage2010analytical}, and remote sensing \cite{long2017accurate, wang2019unsupervised}---has traditionally relied on increasing aperture size.
While a larger diameter sharpens the diffraction limit \cite{choi2024telescope}, 
large telescopes face scalability challenges due to the nonlinear scaling of manufacturing costs \cite{van2004scaling, stahl2020parametric} and the complexity of synthetic-aperture techniques \cite{monnier2003optical, dravins2015optical}. Achieving higher-resolution imaging without increasing aperture size remains an open problem.

\begin{figure}
\centering\includegraphics[width=0.8\textwidth]{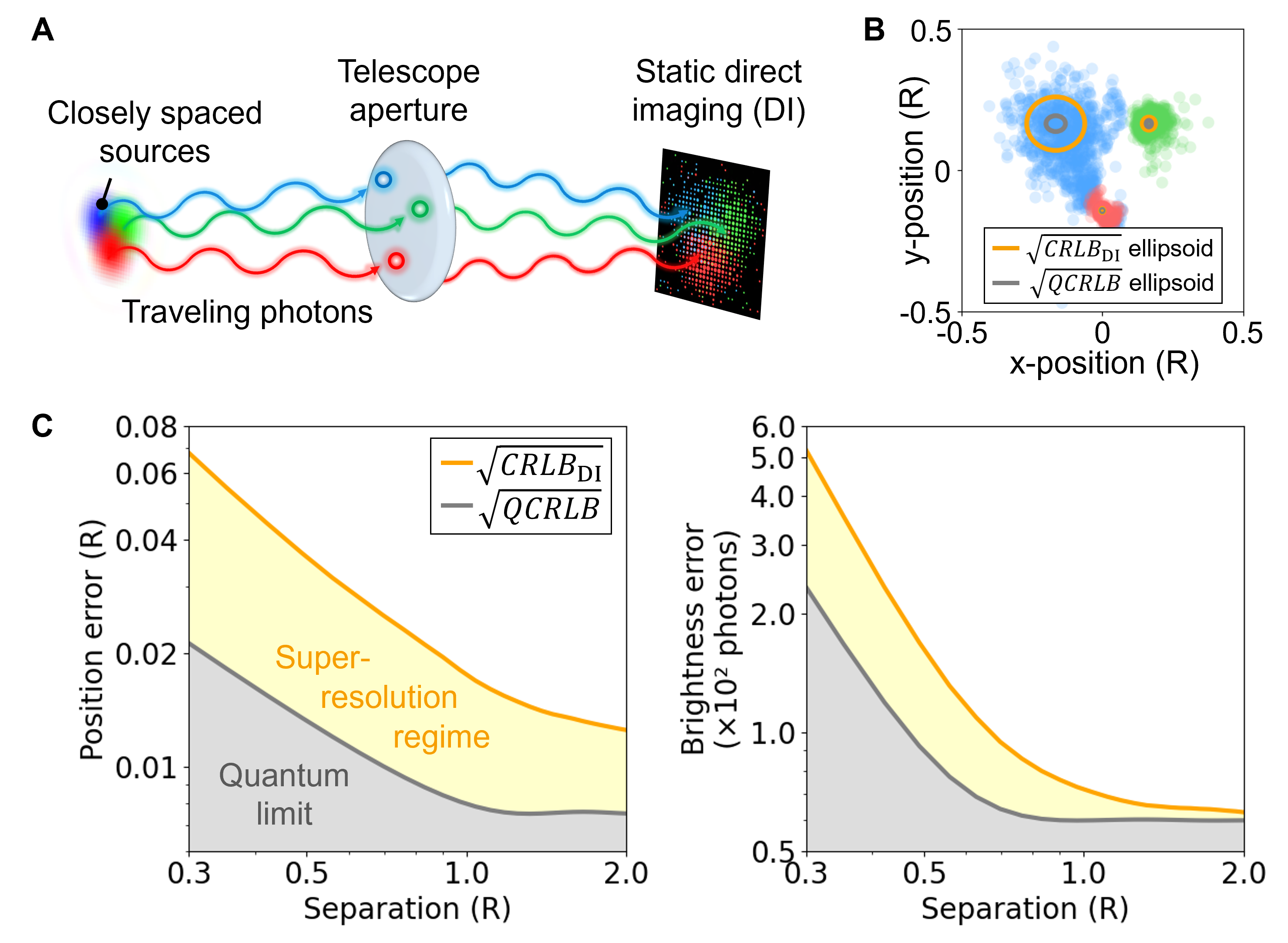}
\caption{
Static direct imaging (DI) and the quantum limit. (A) Passive DI acquisition. (B, C) DI estimation performance for three sources with brightness ratio $[1,5,20]$ (blue, green, and red) at a total photon budget of 14,000. Orange and gray line denote the theoretical performance of DI ($\sqrt{CRLB_\text{DI}}$) and the quantum bound ($\sqrt{QCRLB}$), respectively. (B) Monte Carlo simulations (500 trials) at a separation $0.33R$ showing the distribution of DI estimates. (C) Theoretical precision of mean position and brightness versus source separation, normalized by the Rayleigh limit (R). The orange-shaded region marks the super-resolution regime, where information is lost to passive DI but remains theoretically accessible under optimal measurement, as bounded by the QCRLB.
}
\label{fig1}
\end{figure}

In traditional imaging, also called direct imaging (DI), the Rayleigh curse describes the collapse of localization precision when closely spaced sources are imaged below the diffraction limit. This collapse is not a fundamental physical limit, but a consequence of passive, non-optimal measurement. Conventional DI passively integrates incident photons, which blurs high-frequency spatial information required for precise localization (Fig.~\ref{fig1}A). This information loss becomes severe in the sub-Rayleigh regime, where source separation falls below the diffraction limit, resulting in a substantial increase of estimation uncertainty (Fig.~\ref{fig1}B). 
However, the quantum Cramer-Rao Lower Bound (CRLB), the ultimate quantum precision limit for any physical measurement, reveals that this sensitivity collapse is a measurement deficiency rather than an unavoidable physical constraint \cite{menart2026definition, vrehavcek2017multiparameter, santamaria2024single}. While the DI CRLB grows rapidly as source separation decreases, the quantum CRLB diverges more slowly than the DI limit, identifying extractable information that persists in the deep sub-Rayleigh regime (Fig.~\ref{fig1}B, C). This disparity defines a "Super-resolution regime" where information typically lost by DI remains recoverable through optimal measurement strategies.

Fluorescence microscopy achieves super-resolution by actively controlling emitter dynamics \cite{hell1994breaking, rust2006sub, betzig2006imaging}. By isolating emitters in space or time, these techniques suppress signal overlap at the source. However, such strategies rely on controllable illumination \cite{picariello2025quantum, won2025programmable, zhang2015sub} and emitter dynamics \cite{dertinger2009fast}, making them inapplicable to the uncontrollable sources inherent in astronomy.
Consequently, astronomical super-resolution imaging prioritizes receiver-side transformations. Spatial-mode demultiplexing (SPADE) has emerged as a key framework, projecting the incoming field onto orthogonal spatial bases to theoretically attain the fundamental quantum precision limit \cite{tsang2016quantum, tsang2017subdiffraction}. 
This modal decomposition framework---typically implemented via Multi-Plane Light Converters (MPLCs) \cite{morizur2010programmable, fontaine2019laguerre} or interferometric mode sorters \cite{leach2002measuring}---has demonstrated potential for super-resolution imaging \cite{zhou2019quantum, boucher2020spatial, santamaria2024single, deshler2025experimental}.

Despite these advances, achieving practical quantum-limited performance remains challenging in blind multi-parameter estimation scenarios, because existing imaging systems \cite{chang2026super, bhusal2022smart} typically rely on fixed or pre-designed frameworks. Without prior knowledge of the scene, optimal information extraction must adapt to the unknown source geometry \cite{bao2021quantum, matlin2022imaging, choi2025super}. In practical ground-based astronomy, this challenge is further complicated by atmospheric turbulence, which distorts the optical field before detection and makes passive or fixed bases insufficient for concentrating information toward the quantum limit. This motivates our proposed closed-loop, information-driven measurement paradigm in which the measurement is updated in response to real-time source estimates.

We introduce Adaptive Information Mining (AIM), a closed-loop framework that transforms the imaging architecture into an active information extraction system. AIM bypasses explicit mode decomposition and directly shapes the optical response to maximize information about scene parameters.
This adaptive feedback couples estimation and measurement design: sequential measurements and corresponding estimates inform the subsequent configuration, enabling the system to gradually improve parameter sensitivity. Monte Carlo simulations demonstrate that AIM significantly outperforms passive DI with an identical photon budget, reducing position and brightness errors to levels approaching the fundamental quantum limit. 
Beyond joint position-brightness estimation, AIM also supports blind source-number detection and turbulence-resilient imaging through the same information-theoretic optimization framework. By removing the need for prior scene knowledge while accounting for wavefront distortion, AIM provides a prior-free route toward quantum-limited astronomical imaging in practical scenarios.

\section{Temporally Evolving AIM Architecture}
\begin{figure}
\centering\includegraphics[width=0.95\textwidth]{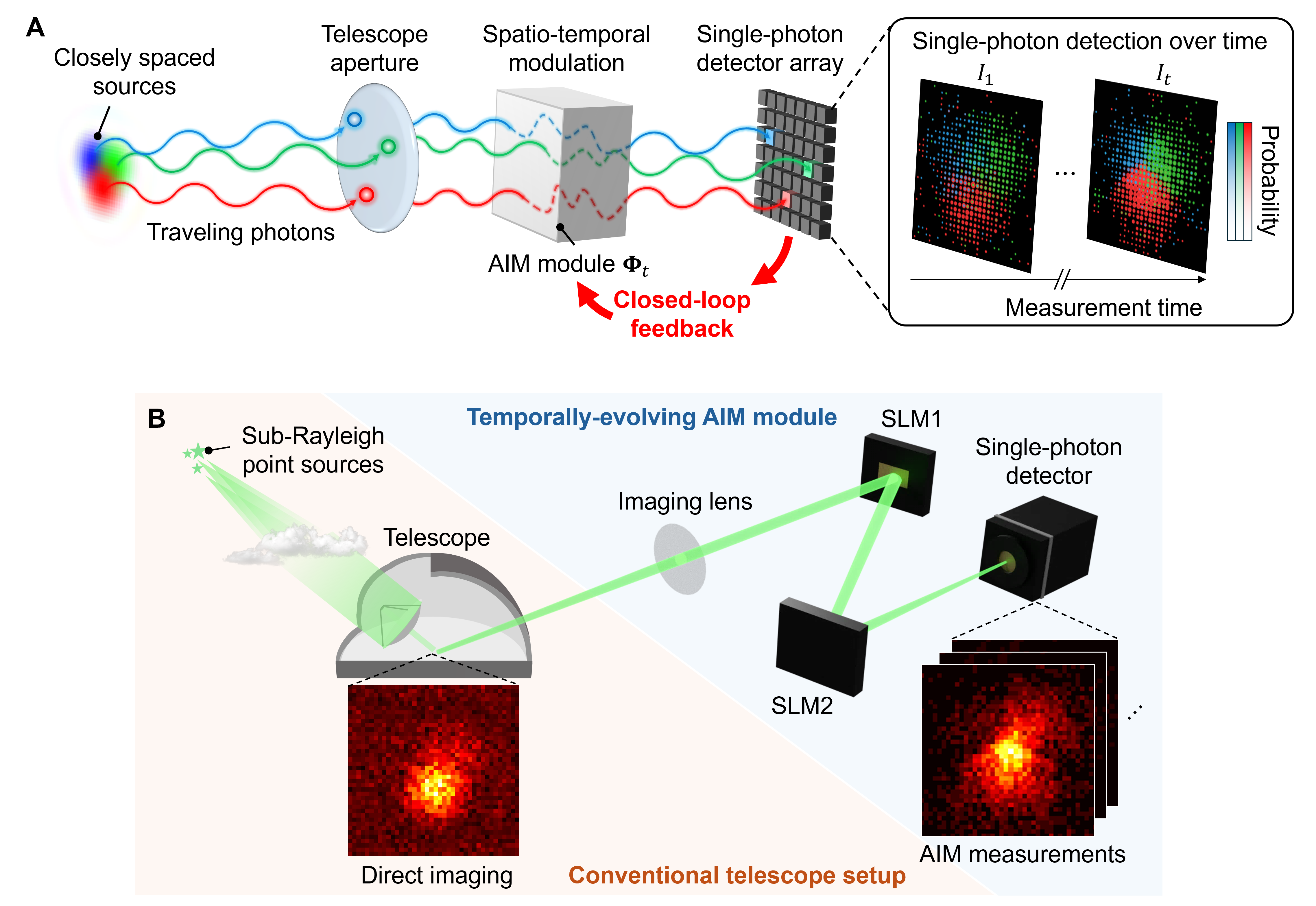}
\caption{
Adaptive Information Mining (AIM) for sub-Rayleigh imaging. (A) AIM operating principle. The AIM module ($\boldsymbol{\Phi}_t$) is adaptively modulated based on prior measurements to drive the subsequent measurement toward high-information states.
(B) Temporally evolving AIM framework compared to conventional direct imaging (DI). Dual SLMs placed after the imaging lens enable adaptive, scene-dependent measurements for resolving sub-Rayleigh point sources.
}
\label{fig2}
\end{figure}

To bridge the gap between classical and quantum limits, we establish a closed-loop AIM framework that transforms imaging from passive integration into active measurement design (Fig.~\ref{fig2}A). 
AIM utilizes a reconfigurable interface based on spatial light modulators (SLMs), which reshapes the optical field based on the previous measurements to generate more informative measurements (Fig.~\ref{fig2}B).

\begin{figure}
\centering\includegraphics[width=0.5\textwidth]{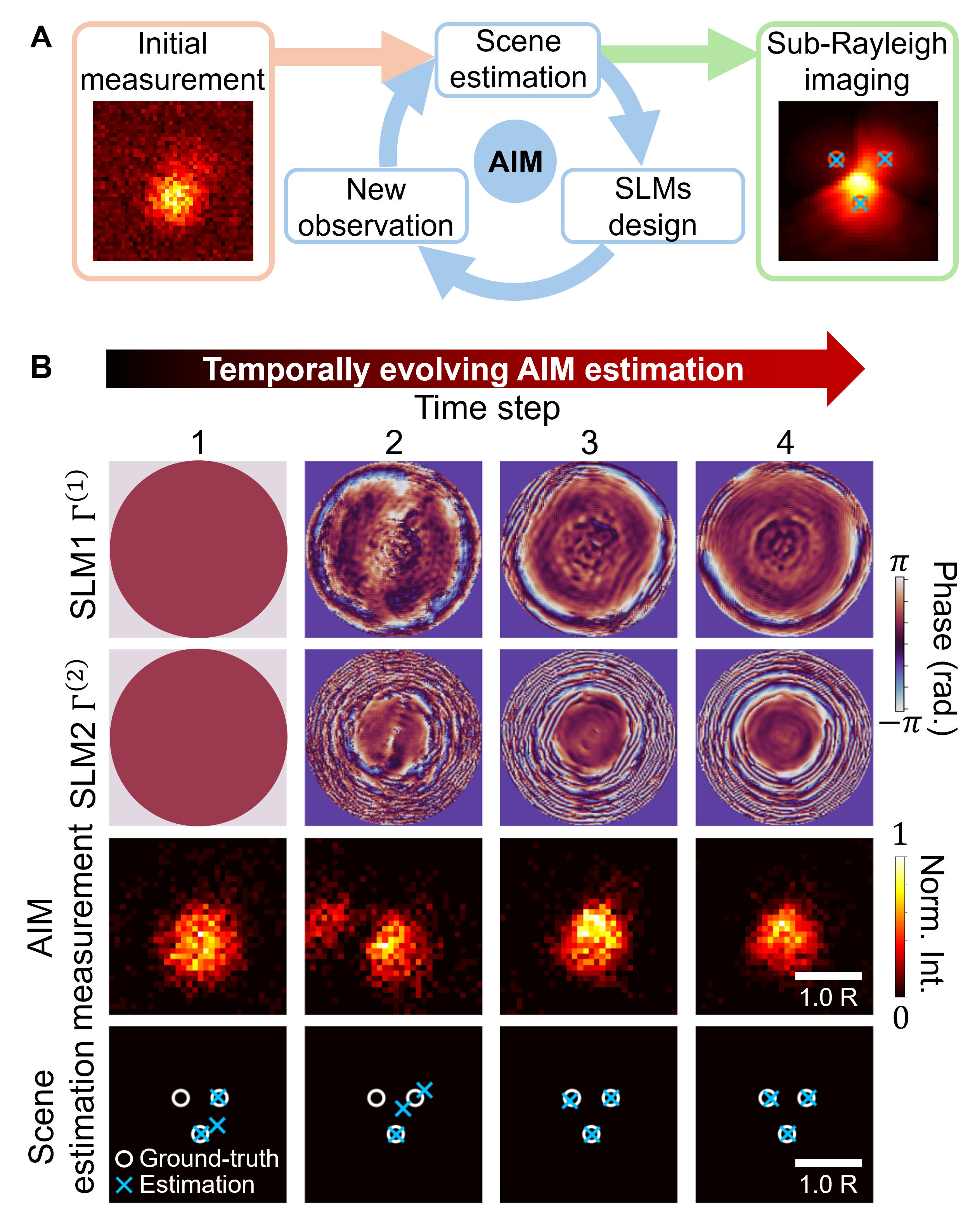}
\caption{
Iterative workflow and temporal evolution of AIM estimation.
(A) Closed-loop AIM workflow. An initial measurement is followed by a three-step iterative cycle---source estimation, information-driven SLM design, and subsequent measurement---enabling sub-Rayleigh imaging. (B) Temporal evolution of AIM estimation. Optimized dual-SLM phase masks $\Gamma^{(1)}$ and $\Gamma^{(2)}$, and corresponding measurements adapt over time to capture high-information states, enabling progressive convergence of estimated source positions (blue crosses) toward the ground truth (white circles).
}
\label{fig3}
\end{figure}

We formulate sub-Rayleigh imaging as an estimation problem. Let the scene be parameterized by $\boldsymbol{\Theta}=\{(x_n, y_n, b_n)\}_{n=1}^{N}\in\mathbb{R}^{3N}$, where $(x_n,y_n)$ and $b_n$ denote the position and brightness of $N$ incoherent point sources. These parameters describe the underlying scene geometry to be inferred from the measurements.
AIM establishes a closed-loop process consisting of three stages: (i) scene estimation, (ii) information-driven SLM pattern design, and (iii) acquisition of a new observation (Fig.~\ref{fig3}A).

In step (i), at each time step $T$, a provisional estimate $\hat{\boldsymbol{\Theta}}_T$ is obtained from the accumulated dataset $\mathcal{D}_T=\{ \boldsymbol{I}_t, \boldsymbol{\Phi}_t \}_{t=1}^{T}$, which includes all prior intensity measurements $\boldsymbol{I}_t$ and their corresponding optical configurations $\boldsymbol{\Phi}_t$. The detection process follows a Poisson model, $\boldsymbol{I}_t \sim \mathrm{Poisson}\!\left(\boldsymbol{\mu}(\boldsymbol{\Theta};\boldsymbol{\Phi}_t)\right)$, and the estimate $\hat{\boldsymbol{\Theta}}_T$ is obtained by minimizing the cumulative negative log-likelihood:
\begin{equation}
\hat{\boldsymbol{\Theta}}_T
=
\arg\min_{\boldsymbol{\Theta}}
\mathcal{L}\!\left(
\boldsymbol{\Theta}
\mid
\mathcal{D}_T
\right)
=
\sum_{t=1}^{T}
\left[
\mathbf{1}^{\!\top}\boldsymbol{\mu}(\boldsymbol{\Theta};\boldsymbol{\Phi}_t)
-
\boldsymbol{I}_t^{\!\top}\log \boldsymbol{\mu}(\boldsymbol{\Theta};\boldsymbol{\Phi}_t)
\right],
\label{eq1.likelihood}
\end{equation}
where $\boldsymbol{\mu}(\boldsymbol{\Theta};\boldsymbol{\Phi}_t) \in \mathbb{R}^{D}$ denotes the photon counts predicted by the forward model across $D$ detector pixels (See appendix), and $\mathbf{1} \in \mathbb{R}^{D}$ is an all-ones vector. This step accumulates information across measurements to refine the current estimate. 

In step (ii), AIM updates the SLM phase configuration to reduce expected uncertainty of $\hat{\boldsymbol{\Theta}}_T$. 
While the AIM framework generalizes to an arbitrary number of SLMs, we employ a two SLM configuration ($\boldsymbol{\Phi}_t=\{\Gamma_t^{(1)},\Gamma_t^{(2)}\}$) for this work, with an analysis of SLM count provided in the Supplementary Material. Local sensitivity is quantified by the classical Fisher Information Matrix (FIM), $\mathcal{J} \in \mathbb{R}^{P \times P}$, evaluated at the current estimate $\hat{\boldsymbol{\mu}} = \boldsymbol{\mu}(\boldsymbol{\Phi}; \hat{\boldsymbol{\Theta}}_T)$:
\begin{equation}
\mathcal{J}(\boldsymbol{\Phi})
=
\left(
{\frac{\partial {\boldsymbol{\mu}}}{\partial \boldsymbol{\Theta}}}\bigg|_{\boldsymbol{\hat{\Theta}}_T}
\right)^{\!\top}
\mathrm{diag}\left(\frac{1}{\hat{\boldsymbol{\mu}}}\right)
\left(
{\frac{\partial {\boldsymbol{\mu}}}{\partial \boldsymbol{\Theta}}}\bigg|_{\boldsymbol{\hat{\Theta}}_T}
\right)
\label{eq:FIM}
\end{equation}
where the Jacobian $\left(
{\frac{\partial {\boldsymbol{\mu}}}{\partial \boldsymbol{\Theta}}}\bigg|_{\boldsymbol{\hat{\Theta}}_T}
\right) \in \mathbb{R}^{D\times P}$ describes how perturbations in the scene parameter affect the expected intensity. The subsequent measurement is then selected to maximize the information content via D-optimality (see Supplementary Material), equivalently minimizing the volume of the associated CRLB uncertainty ellipsoid for all parameters:
\begin{equation}
\boldsymbol{\Phi}^*_{T+1}
= \arg \min_{\boldsymbol{\Phi}} \left[\det \left( \mathcal{J}^{-1}(\boldsymbol{\Phi};\hat{\boldsymbol{\Theta}}_T)\right) \right]
\end{equation}
Within each AIM iteration, steps (i) and (ii) are iteratively optimized using 1,000 and 200 Adam updates, respectively \cite{kingma2014adam}.


At the final step (iii), the updated configuration  $\boldsymbol{\Phi}^*_{T+1}$ is applied to the SLMs, and a new observation $\boldsymbol{I}_{T+1}$ is acquired. This measurement is appended to the dataset, $\mathcal{D}_{T+1}=\mathcal{D}_T\cup \{ \boldsymbol{I}_{T+1}, \boldsymbol{\Phi}^*_{T+1} \}$, completing the cycle. Through this process, AIM generates measurements that are tailored to the scene being imaged.

Fig.~\ref{fig3}B illustrates this temporal evolution of AIM. The SLM patterns are initially flat, corresponding to conventional DI imaging, and become increasingly structured as AIM learns the scene. The resulting measurements become progressively more informative, allowing the estimated source positions (blue crosses) to gradually converge toward the ground truth (white circles).
Importantly, the provisional estimate $\hat{\boldsymbol{\Theta}}_T$ serves as a guide for measurement design rather than a final decision. Early inaccuracies are naturally corrected, as each subsequent measurement is configured to probe the current estimate. 

\section{Scene-Specific AIM Spatial Response}
\begin{figure}
\centering\includegraphics[width=0.70\textwidth]{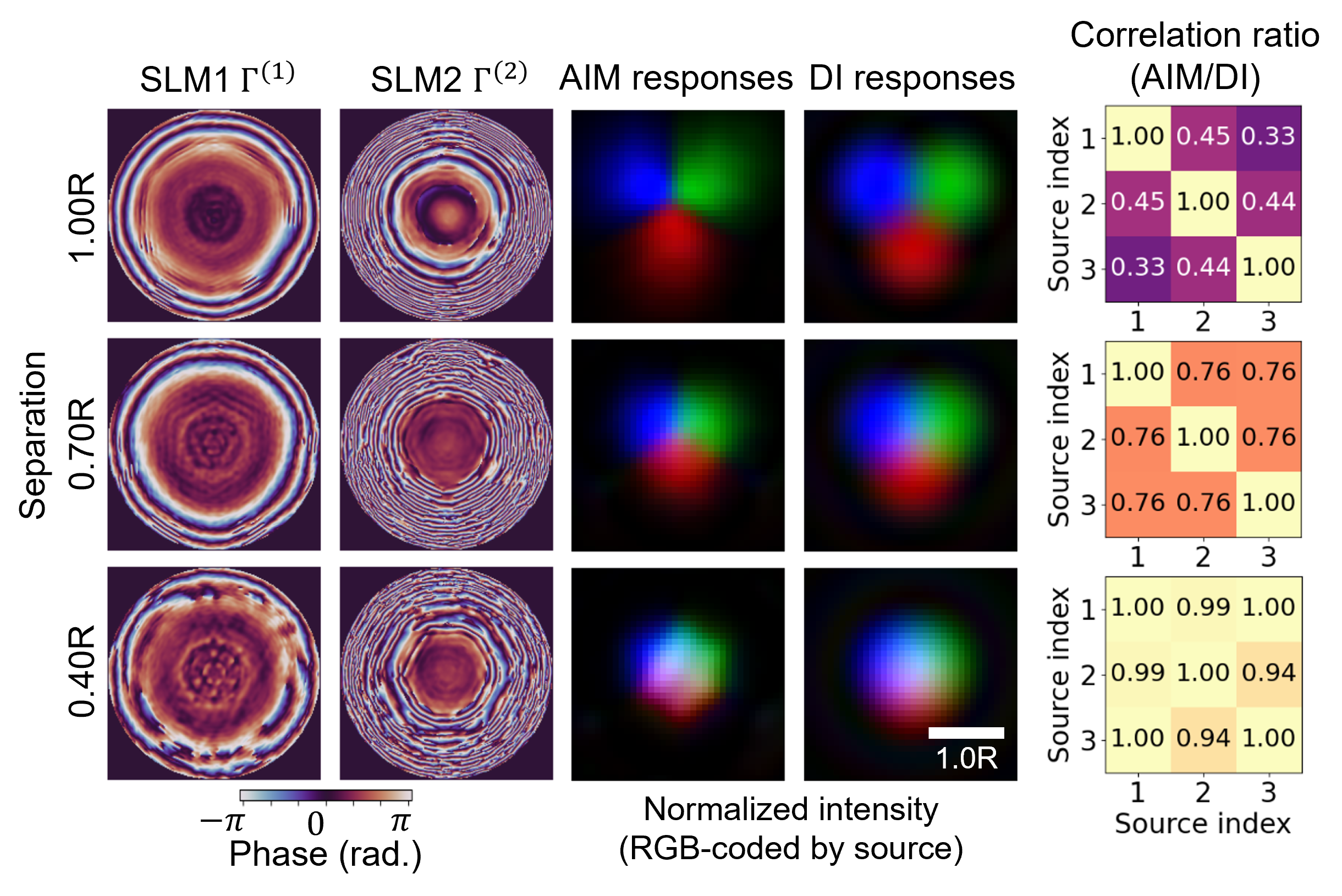}
\caption{
Inter-source overlap suppression via AIM for a three point source scene with brightness ratios [1: 3: 6]. Columns 1-2: optimized SLM phase patterns after convergence for different separations.
Columns 3-4: source-resolved responses under AIM and DI (color-coded).
Columns 5: inter-source correlation matrices, where off-diagonal elements quantify inter-source correlation, highlighting the overlap reduction achieved by AIM.
}
\label{fig4}
\end{figure}

As AIM updates the scene estimate, it redesigns the SLM modulation for that estimated scene, producing a scene-specific spatial response. This response maps closely spaced point sources onto source-specific intensity patterns. With two SLMs placed downstream of the imaging lens (Fig.~\ref{fig2}B), light from different point sources follows distinct modulation paths, which enables these source-specific responses and helps overcome sub-Rayleigh ambiguity.

Fig.~\ref{fig4} visualizes these responses at convergence for different separations. 
To evaluate the overlapping behavior, we compute the spatial correlation between the source-resolved responses, defined as $\rho_{ij}=\frac{\langle \mathbf{s}_i,\mathbf{s}_j\rangle}
{\|\mathbf{s}_i\|_2\,\|\mathbf{s}_j\|_2}$, where $\mathbf{s}_i \in \mathbb{R}^D$ is the vectorized response of the $i$-th source.
The resulting correlation matrices (Fig.~\ref{fig4}, rightmost column) show that, at the Rayleigh-limit separation of $1.0R$, AIM suppresses inter-source correlations more effectively than DI. By making the measurement responses from different sources more distinguishable, AIM reduces estimation ambiguity, providing a clear geometric explanation for its improved accuracy.
At the deep sub-Rayleigh separation of $0.4R$, however, the correlation ratio of AIM rises and approaches that of DI, revealing a shift in the mechanism of information gain. While reduced correlation explains the advantage of AIM at $1.0R$, its performance at $0.4R$ is driven by point source dependent structural variations that cannot be fully captured by a simple correlation metric. 
This behavior underscores that AIM does not only seek spatial decorrelation. Instead, it optimizes the measurement to maximize Fisher information, preserving parameter sensitivity through point source dependent measurement structures even when the source responses remain highly correlated.

\section{Sub-Rayleigh Imaging Performance}
\begin{figure}
\centering\includegraphics[width=0.8\textwidth]{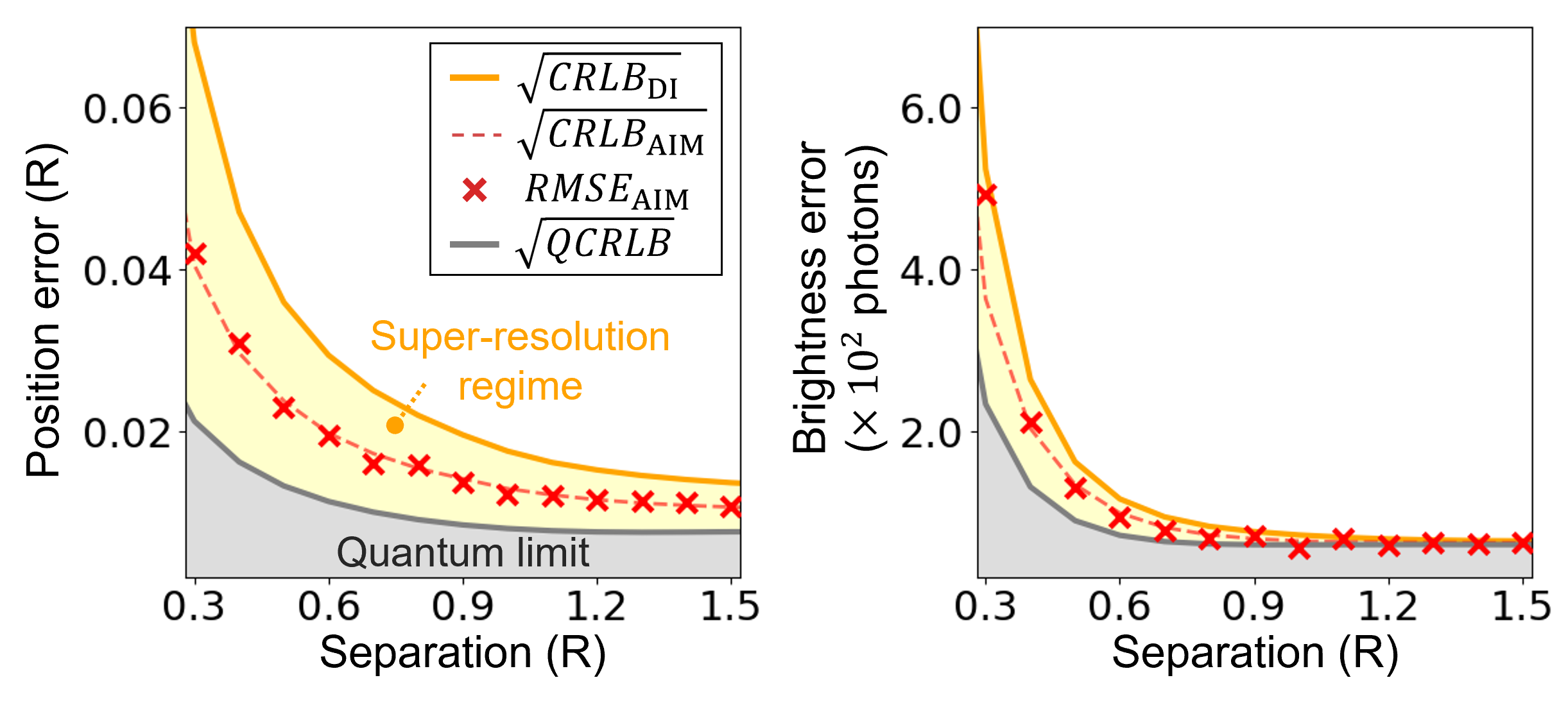}
\caption{Super-resolution performance of AIM using five sequential measurements, with position and brightness RMSE plotted as a function of source separation. Errors are averaged across three sources from 50 Monte Carlo trials per condition (Red cross). Yellow and red curves represent the classical CRLB for DI and AIM, respectively, while the gray curves denote the fundamental quantum CRLB.
All results correspond to brightness ratios [1: 5: 20] and a total photon budget of 14,000. The total photon budget is equally distributed across each AIM measurement.
}
\label{fig5}
\end{figure}
To validate AIM's sub-Rayleigh imaging performance, we conducted Monte Carlo simulations for a three point source scene with unequal brightness ratios (Fig.~\ref{fig5}). To assess performance, the root-mean-squared error (RMSE) and corresponding $\sqrt{\text{CRLB}}$ of AIM were averaged over 50 independent trials for each condition and across all point sources. Note that the performance of AIM reflects the total information gathered across all accumulated measurements.
We evaluate AIM against two benchmarks: the classical DI limit ($\sqrt{\text{CRLB}_{\text{DI}}}$) and the ultimate quantum precision limit ($\sqrt{\text{QCRLB}}$) (see Appendix for QCRLB calculation).
In multi-parameter scenarios with unequal brightness, both $\sqrt{\text{CRLB}_{\text{DI}}}$ and $\sqrt{\text{QCRLB}}$ diverge as separation vanishes---unlike the equal-brightness case where the quantum limit remains finite \cite{vrehavcek2017multiparameter, santamaria2024single}. However, the remaining substantial gap between classical and quantum performance reveals additional extractable information. 
Recall that the super-resolution regime (shaded yellow) marks the gap between these two bounds.

Quantitative analysis of both position and brightness estimation shows that $\text{RMSE}_{\text{AIM}}$ (red crosses) consistently outperforms the classical $\sqrt{\text{CRLB}_{\text{DI}}}$. While the DI performance degrades sharply at small separations, AIM maintains a stable error trajectory that more closely tracks the $\sqrt{\text{QCRLB}}$ even down to $0.3R$. 
To quantify this improvement relative to the quantum floor, we evaluate the error reduction ratio, defined as: $(\sqrt{CRLB_\text{DI}}-\sqrt{QCRLB})/(\sqrt{CRLB_\text{AIM}}-\sqrt{QCRLB})$. This metric shows that AIM's estimation error is around 2--3$\times$ closer to the quantum limit for position and brightness (See Supplementary Information). This proximity to the quantum limit signifies a transition to information-optimal imaging, where AIM’s adaptive, spatially-varying response suppresses the estimation error. Further performance gains with increasing number of SLMs are provided in Supplementary Information.

Beyond parameter precision, AIM improves source number detection probability. This advantage is particularly critical in fully blind scenarios without prior knowledge of the number of point sources \cite{deshler2026quantum}. Detailed quantitative evaluations of the detection probability are provided in the Supplementary Information.

\section{Single-parameter targeted AIM design}
\begin{figure}
\centering\includegraphics[width=0.5\textwidth]{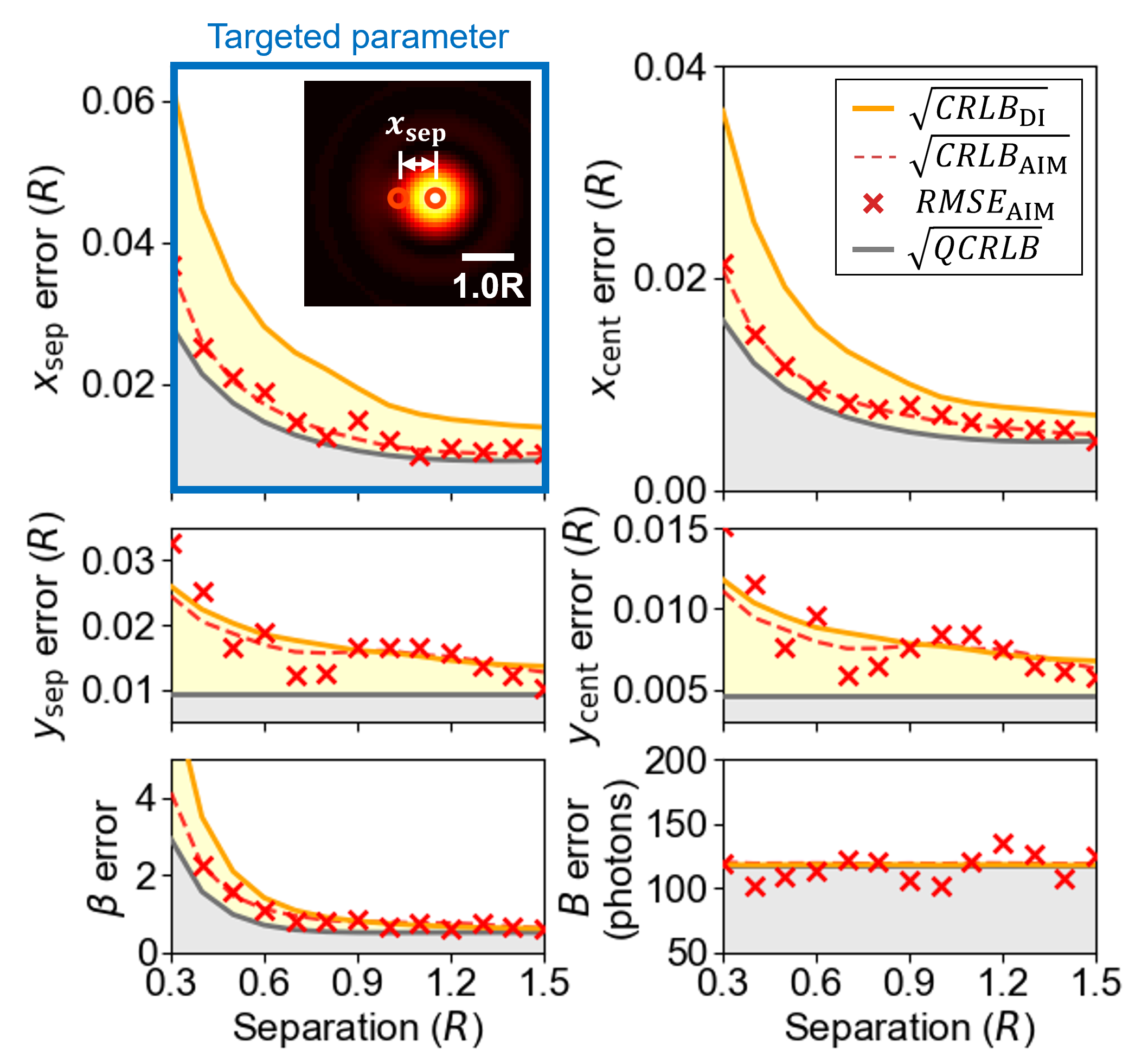}
\caption{Single-parameter targeted AIM performance for a two-point-source scene using five sequential measurements. The two unequal-brightness sources are separated primarily along the $x$ direction, and the RMSE of each reparameterized scene parameter is plotted as a function of source separation. The SLM pattern is optimized for the marginalized CRLB of the target parameter $x_{\rm sep}$ while all parameters are jointly estimated. Scatter points denote RMSEs averaged over 50 independent trials per condition. Yellow and red curves represent the classical CRLB for DI and AIM, respectively, while the gray curves denote the fundamental quantum CRLB. All results correspond to brightness ratios [1:15] and a total photon budget of 14,000, equally distributed across each AIM measurement.
}
\label{fig6}
\end{figure}
Although AIM is optimized to increase information across the full set of unknown scene parameters, certain imaging tasks place greater emphasis on specific scene parameters. In binary-source super-resolution, for example, the source separation is often of particular interest \cite{tsang2016quantum}. AIM naturally accommodates such task-specific priorities by tailoring the measurement design specifically for a chosen target parameter while still jointly estimating all other unknown scene parameters.

To demonstrate this single-parameter targeted optimization, we consider a two-point-sources scene with unequal brightness. Following the relative-coordinate parametrization commonly used in binary super-resolution analysis, we map the absolute scene parameters $\boldsymbol{\Theta} = \{(x_n, y_n, b_n)\}_{n=1}^2$ to a relative parameter space:
\begin{equation}
\boldsymbol{\Theta}_{\rm rel} = [y_{\rm sep}, x_{\rm sep}, \beta, y_{\rm cent}, x_{\rm cent}, B]^T,
\end{equation}
where $y_{\rm sep}=y_2-y_1$ and $x_{\rm sep}=x_2-x_1$ denote the spatial separations, $\beta=b_2/b_1$ is the brightness constrast ratio, $y_{\rm cent}=(y_1+y_2)/2$ and $x_{\rm cent}=(x_1+x_2)/2$ represent the centroid positions, and $B=b_1+b_2$ is the total photon budget. Here, the separation along the primary axis, $x_{\rm sep}$, is selected as the target parameter of interest, while the remaining five variables act as unknown nuisance parameters. During the measurement-design step, instead of applying D-optimality to the full FIM, the SLM patterns are driven to minimize only the marginalized CRLB of $x_{\rm sep}$ (i.e., the corresponding diagonal element of the inverse multi-parameter FIM). This objective concentrates the subsequent measurement on information extraction for the target parameter $x_{\rm sep}$, while accounting for uncertainty and coupling with the nuisance parameters.

Fig.~\ref{fig6} demonstrates how single-parameter optimization enables AIM to extract the target parameter near the fundamental quantum limit with multiple unknown nuisance parameters. For the targeted parameter $x_{\rm sep}$, both the AIM CRLB and the empirical reconstruction RMSE closely approach the QCRLB across the entire separation range. This targeted design achieves an estimation error approximately 5$\times$ closer to the quantum limit than DI. Crucially, despite the measurement optimization being dedicated solely to $x_{\rm sep}$, the non-targeted nuisance parameters remain well-conditioned, yielding errors comparable to or lower than the DI benchmark. These findings demonstrate that AIM can either balance performance across all parameters or selectively mine information for more important parameters of interest to approach quantum-limited performance more closely.

\section{Turbulence-resilient imaging}
\begin{figure}
\centering\includegraphics[width=1.0\textwidth]{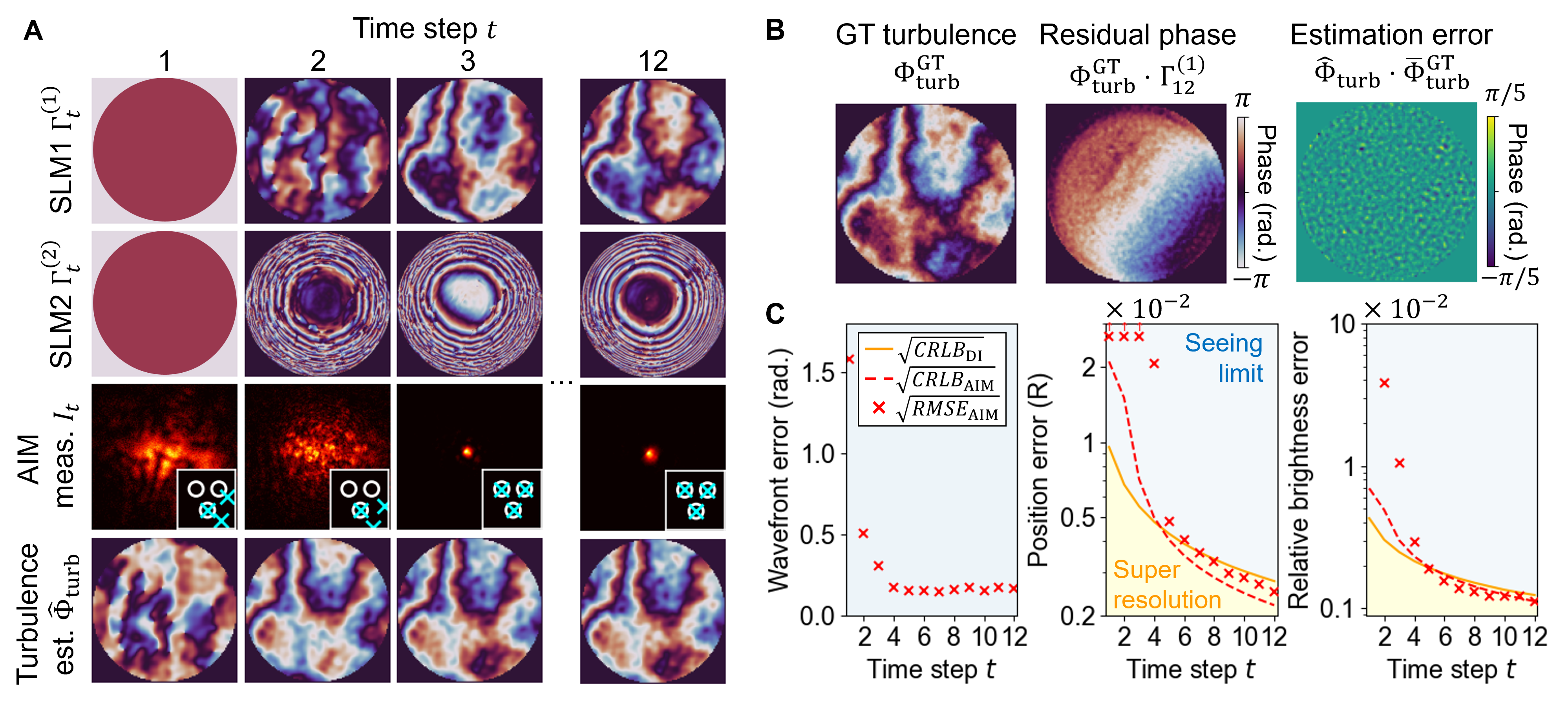}
\caption{Turbulence-resilient AIM with joint turbulent wavefront estimation. (A) AIM evolution under fixed turbulence ($D/r_0=8$). $\Gamma^{(1)}$ is placed at the pupil plane and $\Gamma^{(2)}$ is placed between the lens and detector.
Rows show the optimized dual-SLM phase masks, AIM measurements with scene estimates, and estimated turbulence phases for the AIM iterations. 
In the insets, blue crosses and white circles denote estimated and ground-truth source positions, respectively. 
(B) Ground-truth turbulence phase, residual phase at the pupil-plane and final turbulence-estimation error.
(C) AIM performance over iterative joint turbulence and scene estimation. turbulence-phase, Position and relative brightness errors are shown versus time step t. 
The scene contains three point sources with separation 0.66 R, brightness ratio [1: 5: 15], and a total photon budget of $1\times10^6$ over 12 AIM measurements, with turbulence strength $D/r_0=8$ and Monte Carlo averaging over 50 independent turbulence realizations.
Relative brightness errors are reported because the absorbing boundary used to suppress periodic-boundary artifacts in Fourier-domain propagation alters the absolute photon throughput, especially under turbulence.
}
\label{fig7}
\end{figure}

The AIM framework can be extended to ground-based astronomical imaging by incorporating atmospheric turbulence into the forward model \cite{choi2026poisson, choi2025photon}. Under a frozen-turbulence condition, where the sequential AIM measurements are assumed to be sufficiently fast, the turbulent phase remains fixed throughout the loop. We estimate the augmented parameter set 
$\boldsymbol{\Theta}_{\mathrm{total}}=\{\boldsymbol{\Theta},\Phi_{\mathrm{turb}}\}$, 
where the estimation step (i) jointly updates the point source parameters $\boldsymbol{\Theta}$ and the turbulent wavefront $\Phi_{\mathrm{turb}}$.
A total variation (TV) prior is used to regularize the wavefront by penalizing its spatial gradients. In contrast, the measurement design step (ii) optimizes the SLM patterns for information gain in the scene parameters $\boldsymbol{\Theta}$. The AIM CRLBs reported here follow the same convention: they bound the scene parameters only, with the turbulent wavefront parameter excluded from the bound calculation.

Fig. \ref{fig7}A shows a representative AIM sequence using a dual-SLM architecture, with $\Gamma^{(1)}$ placed at the aperture pupil plane and $\Gamma^{(2)}$ placed after the imaging lens; details of the optical setup are provided in the Supplementary Information. Initially, both SLMs are flat, producing a turbulence-degraded speckle measurement. Over successive AIM iterations, SLM1 $\Gamma^{(1)}$ evolves toward a pupil-plane modulation that counteracts the turbulent phase, while SLM2 $\Gamma^{(2)}$ forms source-specific responses. The resulting measurements evolve from diffuse speckle patterns into localized responses, and the updated scene and turbulence estimates progressively converge.

Fig.~\ref{fig7}B examines the estimated turbulent phase. The recovered turbulent phase closely matches the ground truth. After applying $\Gamma^{(1)}$, the residual pupil-plane phase is substantially reduced and is dominated by low-order components such as tip/tilt and weak defocus. Although AIM does not explicitly optimize a wavefront-flattening objective, the scene-information objective naturally produces an adaptive optics (AO)-like pupil plane correction.

Fig.~\ref{fig7}C quantifies the performance over 12 sequential AIM measurements, averaged over 50 independent turbulence realizations at $D/r_0=8$. The phase panel shows rapid convergence of the turbulence estimate, with the phase RMSE saturating at approximately 0.2 rad, corresponding to about $0.03\lambda$ wavefront error. The position and brightness panels compare the AIM reconstruction RMSEs (red crosses), accumulated AIM CRLBs (red dashed curves), and diffraction-limited DI CRLBs (orange curves). As the wavefront error decreases, the scene RMSEs closely track the AIM CRLBs, enabling the scene error to fall below the diffraction-limited DI benchmark. The slight RMSE--CRLB gap especially in position error is expected because the estimator jointly recovers the unknown turbulent wavefront, while the reported CRLBs quantify only the scene-parameter information.

Overall, Fig.~\ref{fig7} demonstrates that AIM combines AO-like pupil plane correction with source-specific measurement design, enabling super-resolution scene recovery beyond the diffraction-limited DI benchmark despite turbulence-induced ambiguity.



\section{Discussion}
AIM achieves performance close to the fundamental quantum limit without relying on explicit mode decomposition, instead directly optimizing the classical Fisher information. This distinguishes it from mode sorting based implementations, where physical constraints such as mode dependent loss and residual crosstalk can degrade performance \cite{chen2024multi, fang2021performance}. In contrast, AIM preserves the available information by embedding the optimization directly into the measurement process.

Our AIM framework naturally extends to complex, multi-parameter scenes. Unlike prior approaches that assume fixed source counts or known brightness ratios \cite{zhou2019quantum, boucher2020spatial, santamaria2024single}, AIM adapts to the intrinsic information structure of the scene, enabling joint estimation of both parameters and source number without prior knowledge. 

Practically, AIM offer a path for ground-based imaging under astronomical turbulence. Unlike conventional systems that require explicit wavefront sensing and AO hardware, AIM treats turbulence as an unknown parameter of the forward model. By maximizing scene information, the AIM naturally evolves its optical response to compensate for wavefront distortion. This information-driven sensing serves as an alternative to standard AO paradigms, maintaining sub-Rayleigh precision even under atmospheric degradation.

Transitioning this sequential AIM framework to real-time ground-based operations requires hardware capable of tracking dynamic atmospheric fluctuations. This temporal requirement can be met using advanced kHz-class SLMs \cite{deng_diffraction_2022,chao_high-brightness_2023} paired with faster quantum detectors.

In summary, AIM establishes a hardware-software co-design paradigm in which optical measurements are dynamically tailored to maximize information extraction. By transforming the imaging architecture into a reconfigurable information mining system, our approach overcomes the limitations of the passive, non-adaptive architectures and enables photon-efficient super-resolution imaging without prior knowledge of the scene or aberrations. This framework provides a scalable foundation for a wide range of imaging paradigms, from deep-space astrometry and exoplanet detection to high-resolution optical microscopy.


\bigskip
\section*{Appendix: Quantum Fisher Information Analysis}

To evaluate the fundamental precision limits of the multi-source imaging system, we derive the multiparameter Quantum Fisher Information Matrix (Q-FIM). This framework quantifies the maximum extractable information for the simultaneous estimation of $3N$ parameters $\boldsymbol{\Theta}=\{(x_n, y_n, b_n)\}_{n=1}^{N}$.

In incoherent imaging under the weak-source approximation \cite{tsang2016quantum, tsang2017subdiffraction, menart2026definition}, the optical field in each temporal mode is restricted to the vacuum and one-photon states. Let $p_n$ denote the one-photon probability contributed by the $n$-th source within a single temporal mode, and let $\epsilon = \sum_{n=1}^N p_n \ll 1$ denote the total one photon probability. The full density matrix $\rho$ is given by
\begin{equation}
\rho \approx (1-\epsilon)|0\rangle\langle 0| + \sum_{n=1}^{N} p_n |\psi_n\rangle \langle \psi_n|,
\tag{A1}
\end{equation}
where $|0\rangle$ denotes the vacuum state, $|\psi_n\rangle \in \mathbb{C}^{D \times 1}$ is the one-photon state emitted by the $n$-th source, and $D$ is the number of detector pixels.
For an observation window comprising $M$ independent temporal modes, the accumulated brightness is defined as $b_n = M p_n$.

To avoid computationally expensive operations in the high-dimensional $D$ pixel space, we identify an augmented spatial subspace that spans all parametric variations. We construct the matrix $\mathbf{V}$ consisting of the derivatives of the fields with respect to all parameters:
\begin{equation}
\mathbf{V} = \left[ \frac{\partial |\psi_1\rangle}{\partial y_1}, \frac{\partial |\psi_1\rangle}{\partial x_1}, |\psi_1\rangle, \dots, \frac{\partial |\psi_N\rangle}{\partial y_N}, \frac{\partial |\psi_N\rangle}{\partial x_N},|\psi_N\rangle\right] \in \mathbb{C}^{D \times 3N}.
\tag{A2}
\end{equation}

Using the Gram matrix $G = \mathbf{V}^\dagger \mathbf{V} \in \mathbb{C}^{3N \times 3N}$, we perform an eigen-decomposition $G = \mathbf{T} \mathbf{D} \mathbf{T}^\dagger$ to extract a $3N$-dimensional orthonormal spatial basis $|u_k\rangle \in \mathbb{C}^{D\times1}$:
\begin{equation}
|u_k\rangle = \frac{1}{\sqrt{\eta_{k}}} \mathbf{V} |v_k\rangle, \quad k = 1, \dots, 3N,
\tag{A3}
\end{equation}
where $\eta_k$ are the diagonal elements of $\mathbf{D}$ and $|v_k\rangle$ are the columns of $\mathbf{T}$.
This transformation compresses the $D$-dimensional pixel data into a $3N$-dimensional minimal operational space.

While $\{|u_k\rangle\}$ spans the necessary spatial support, the density matrix $\rho$ is generally not diagonal in this representation. To solve the Symmetric Logarithmic Derivative (SLD) equation, we must transition the system into the eigenbasis of the density matrix $\rho$. 
We project the nontrivial one-photon component $\rho_1$ into this subspace:
\begin{equation}
[\rho_1]_{ij} = \sum_{n=1}^{N} p_n \langle u_i | \psi_n \rangle \langle \psi_n | u_j \rangle.
\tag{A4}
\end{equation}

By diagonalizing this reduced matrix as $[\rho_1]\mathbf{c}_m=\lambda_m \mathbf{c}_m$, we obtain the eigenvalues $\lambda_m$ and the corresponding eigenvectors $\mathbf{c}_m$. These coefficients define the physical one-photon eigenmodes $|e_m\rangle \in \mathbb{C}^{D \times 1}$:
\begin{equation}
|e_m\rangle = \sum_{k=1}^{3N} c_{m,k} |u_k\rangle.
\tag{A5}
\end{equation}

Due to the orthogonality of the vacuum and the need to track the total probability conservation for absolute brightness estimation, we explicitly incorporate the vacuum mode into our operational basis. We define the extended set of eigenmodes $\{|e_m\rangle\}_{m=0}^{3N}$, where $|e_0\rangle = |0\rangle$ represents the vacuum mode with the eigenvalue $\lambda_0 = 1 - \epsilon$, and the remaining modes ($m \ge 1$) correspond to the one-photon eigenmodes with eigenvalues $\lambda_m$.

In this full representation, the density matrix is diagonalized as $\rho = \sum_{m=0}^{3N} \lambda_m |e_m\rangle \langle e_m|$. The SLD operator $L_\mu$ for each parameter $\theta_\mu \in \Theta$ is determined by solving
\begin{equation}
\langle e_m| L_\mu|e_l \rangle=\frac{2}{\lambda_m +\lambda_l} \langle e_m|\partial_\mu\rho|e_l\rangle, \quad m,l \in \{0, 1, \dots, 3N\},
\tag{A6}
\label{eq:B6}
\end{equation}
where $\partial_\mu \rho$ is the derivative of the full density matrix with respect to $\theta_\mu$.

Let $\alpha_{m,n} = \langle e_m | \psi_n \rangle$ and $\beta_{m,\mu} = \langle e_m | \partial_\mu \psi_n \rangle$ be the projection of the fields and their derivatives onto the eigenmodes (noting that $\alpha_{0,n}=0$ and $\beta_{0,\mu}=0$ due to orthogonality with the vacuum). For a parameter $\theta_\mu$ associated with the $n$-th source, the matrix elements of $\partial_\mu \rho$ are given by
\begin{equation}
\langle e_m | \partial_\mu \rho | e_l \rangle =
\begin{cases}
\frac{b_n}{M} \left( \beta_{m,\mu} \alpha_{l,n}^* + \alpha_{m,n} \beta_{l,\mu}^* \right), & \text{for } \theta_\mu \in \{y_n, x_n\} \text{ and } m,l \ge 1, \\
\frac{1}{M} \left( \alpha_{m,n} \alpha_{l,n}^* - \delta_{m,0}\delta_{l,0} \right), & \text{for } \theta_\mu = b_n.
\end{cases}
\tag{A7}
\label{eq:B7}
\end{equation}
where the Kronecker delta $\delta_{m,0}\delta_{l,0}$ correctly accounts for the penalty in the vacuum population as the absolute brightness of a source increases.
Using these matrix elements, the single-mode Q-FIM is then computed as
\begin{equation}
\mathcal{H}^{(1)}_{\mu\nu} = \mathrm{Re} \left[ \sum_{m=0}^{3N} \lambda_m \langle e_m | L_\mu L_\nu | e_m \rangle \right].
\tag{A8}
\end{equation}

For multiparameter quantum estimation, the SLD QCRLB is not generally guaranteed to be simultaneously attainable because the optimal measurements for different parameters may be incompatible. For the incoherent far-field imaging model considered here, we assume the weak-commutativity condition \cite{vrehavcek2017multiparameter, sidhu2021tight}:
\begin{equation}
\mathrm{Im} \{\mathrm{Tr}\left[\rho L_\mu L_\nu\right] \}=0,\qquad \forall \mu,\nu ,
\tag{A9}
\end{equation}
for the source-position and brightness parameters used in our Q-FIM calculations. When this condition holds, the fundamental multi-parameter Holevo bound coincides with the SLD-QCRLB \cite{sidhu2021tight, albarelli2019evaluating}. In such compatible models, the SLD-QCRLB can be regarded as the asymptotically attainable multiparameter quantum benchmark, in principle reachable by collective measurements over many independent copies \cite{conlon2021efficient}.

For an observation window comprising $M$ independent temporal modes, the total Q-FIM is additive across all spatial and brightness parameters:
\begin{equation}
\mathcal{H}_{\mu\nu} = M\, \mathcal{H}^{(1)}_{\mu\nu},
\tag{A10}
\end{equation}
Notably, the exact single-source entry of the derived Q-FIM evaluated with the 0-photon (vacuum) state yields $\mathcal{H}_{bb} = 1 / [b(1 - b/M)]$, whereas omitting the vacuum state yields $\mathcal{H}_{bb} = 1/b$. In the asymptotic limit where the observation window is sufficiently large ($b/M \to 0$), the two formulations converge identically as $\mathcal{H}_{bb} \to 1/b$. Considering this asymptotic equivalence and to avoid unnecessary computational overhead from high-dimensional vacuum padding, we neglect the 0-photon state in our practical evaluation of Eq.~\ref{eq:B7}, thereby retaining the highly efficient $3N$-dimensional operational subspace framework.

Consequently, by evaluating the single-mode derivatives via Eq.~\ref{eq:B7} under this regime, the final additive Q-FIM across $M$ temporal modes scales as $\mathcal{H}_{\mu\nu} = M \mathcal{H}^{(1)}_{\mu\nu}$. Under the compatibility condition, the inverse matrix $\mathcal{H}^{-1}$ yields the multiparameter SLD Quantum Cramer-Rao Lower Bound used as the quantum benchmark for simultaneous estimation of source positions and absolute intensities.

\bigskip
\begin{backmatter}
\bmsection{Funding}
Air Force Office of Scientific Research (FA9550-24-1-0345). 

\bmsection{Disclosures}
The authors declare no conflicts of interest.

\bmsection{Data Availability}
All data needed to evaluate the conclusions in the paper are present in the paper and/or the Supplementary Materials. 

\bmsection{Supplemental document}
See Supplement 1 for supporting content.

\end{backmatter}


\bibliography{reference_correct}

@article{deshler2025experimental,
  title={Experimental demonstration of a quantum-optimal coronagraph using spatial mode sorters},
  author={Deshler, Nico and Ozer, Itay and Ashok, Amit and Guha, Saikat},
  journal={Optica},
  volume={12},
  number={4},
  pages={518--529},
  year={2025},
  publisher={Optica Publishing Group}
}

@article{conlon2021efficient,
  title={Efficient computation of the Nagaoka--Hayashi bound for multiparameter estimation with separable measurements},
  author={Conlon, Lorc{\'a}n O and Suzuki, Jun and Lam, Ping Koy and Assad, Syed M},
  journal={npj Quantum Information},
  volume={7},
  number={1},
  pages={110},
  year={2021},
  publisher={Nature Publishing Group UK London}
}

@article{albarelli2019evaluating,
  title={Evaluating the Holevo Cram$\backslash$'er-Rao bound for multi-parameter quantum metrology},
  author={Albarelli, Francesco and Friel, Jamie F and Datta, Animesh},
  journal={arXiv preprint arXiv:1906.05724},
  year={2019}
}

@article{sidhu2021tight,
  title={Tight bounds on the simultaneous estimation of incompatible parameters},
  author={Sidhu, Jasminder S and Ouyang, Yingkai and Campbell, Earl T and Kok, Pieter},
  journal={Physical Review X},
  volume={11},
  number={1},
  pages={011028},
  year={2021},
  publisher={APS}
}

@article{deshler2026quantum,
  title={Quantum limits of exoplanet detection and localization},
  author={Deshler, Nico and Haffert, Sebastiaan and Ashok, Amit},
  journal={Physical Review A},
  volume={113},
  number={4},
  pages={042406},
  year={2026},
  publisher={APS}
}

@article{matlin2022imaging,
  title={Imaging arbitrary incoherent source distributions with near quantum-limited resolution},
  author={Matlin, Erik F and Zipp, Lucas J},
  journal={Scientific Reports},
  volume={12},
  number={1},
  pages={2810},
  year={2022},
  publisher={Nature Publishing Group UK London}
}

@article{long2017accurate,
  title={Accurate object localization in remote sensing images based on convolutional neural networks},
  author={Long, Yang and Gong, Yiping and Xiao, Zhifeng and Liu, Qing},
  journal={IEEE Transactions on Geoscience and Remote Sensing},
  volume={55},
  number={5},
  pages={2486--2498},
  year={2017},
  publisher={IEEE}
}

@article{choi2024telescope,
  title={Telescope imaging beyond the Rayleigh limit in extremely low SNR},
  author={Choi, Hyunsoo and Choi, Seungman and Menart, Peter and Deka, Angshuman and Jacob, Zubin},
  journal={New Journal of Physics},
  volume={26},
  number={9},
  pages={093019},
  year={2024},
  publisher={IOP Publishing}
}

@article{bao2021quantum,
  title={Quantum-accelerated imaging of N stars},
  author={Bao, Fanglin and Choi, Hyunsoo and Aggarwal, Vaneet and Jacob, Zubin},
  journal={Optics Letters},
  volume={46},
  number={13},
  pages={3045--3048},
  year={2021},
  publisher={Optical Society of America}
}

@article{sauvage2010analytical,
  title={Analytical expression of long-exposure adaptive-optics-corrected coronagraphic image. First application to exoplanet detection},
  author={Sauvage, J-F and Mugnier, Laurent M and Rousset, G{\'e}rard and Fusco, Thierry},
  journal={Journal of the Optical Society of America A},
  volume={27},
  number={11},
  pages={A157--A170},
  year={2010},
  publisher={OSA}
}

@inproceedings{wang2019unsupervised,
  title={Unsupervised remote sensing image super-resolution using cycle CNN},
  author={Wang, Pengrui and Zhang, Haopeng and Zhou, Feng and Jiang, Zhiguo},
  booktitle={IGARSS 2019-2019 IEEE International Geoscience and Remote Sensing Symposium},
  pages={3117--3120},
  year={2019},
  organization={IEEE}
}

@inproceedings{van2004scaling,
  title={The scaling relationship between telescope cost and aperture size for very large telescopes},
  author={Van Belle, Gerard Theodore and Meinel, Aden Baker and Meinel, Marjorie Pettit},
  booktitle={Ground-based telescopes},
  volume={5489},
  pages={563--570},
  year={2004},
  organization={Spie}
}

@inproceedings{stahl2020parametric,
  title={Parametric cost model for ground and space telescopes},
  author={Stahl, H Philip and Allison, Michael},
  booktitle={Modeling, systems engineering, and project management for astronomy IX},
  volume={11450},
  pages={262--273},
  year={2020},
  organization={SPIE}
}

@article{monnier2003optical,
  title={Optical interferometry in astronomy},
  author={Monnier, John D},
  journal={Reports on Progress in Physics},
  volume={66},
  number={5},
  pages={789--857},
  year={2003}
}

@article{dravins2015optical,
  title={Optical aperture synthesis with electronically connected telescopes},
  author={Dravins, Dainis and Lagadec, Tiphaine and Nu{\~n}ez, Paul D},
  journal={Nature communications},
  volume={6},
  number={1},
  pages={6852},
  year={2015},
  publisher={Nature Publishing Group UK London}
}

@article{tsang2016quantum,
  title={Quantum theory of superresolution for two incoherent optical point sources},
  author={Tsang, Mankei and Nair, Ranjith and Lu, Xiao-Ming},
  journal={Physical Review X},
  volume={6},
  number={3},
  pages={031033},
  year={2016},
  publisher={APS}
}

@article{hell1994breaking,
  title={Breaking the diffraction resolution limit by stimulated emission: stimulated-emission-depletion fluorescence microscopy},
  author={Hell, Stefan W and Wichmann, Jan},
  journal={Optics letters},
  volume={19},
  number={11},
  pages={780--782},
  year={1994},
  publisher={Optical Society of America}
}

@article{rust2006sub,
  title={Sub-diffraction-limit imaging by stochastic optical reconstruction microscopy (STORM)},
  author={Rust, Michael J and Bates, Mark and Zhuang, Xiaowei},
  journal={Nature methods},
  volume={3},
  number={10},
  pages={793--796},
  year={2006},
  publisher={Nature Publishing Group US New York}
}

@article{betzig2006imaging,
  title={Imaging intracellular fluorescent proteins at nanometer resolution},
  author={Betzig, Eric and Patterson, George H and Sougrat, Rachid and Lindwasser, O Wolf and Olenych, Scott and Bonifacino, Juan S and Davidson, Michael W and Lippincott-Schwartz, Jennifer and Hess, Harald F},
  journal={science},
  volume={313},
  number={5793},
  pages={1642--1645},
  year={2006},
  publisher={American Association for the Advancement of Science}
}

@article{picariello2025quantum,
  title={Quantum super-resolution microscopy by photon statistics and structured light},
  author={Picariello, Fabio and Losero, Elena and Ditalia Tchernij, Sviatoslav and Boucher, Pauline and Genovese, Marco and Ruo-Berchera, Ivano and Degiovanni, Ivo Pietro},
  journal={Optica},
  volume={12},
  number={4},
  pages={490--497},
  year={2025},
  publisher={Optica Publishing Group}
}

@article{won2025programmable,
  title={Programmable structured DUV illumination by coherent harmonic generation at crystalline solids for nanometer-resolution inspection of periodic samples},
  author={Won, Seungjai and Kim, Jungyoon and Oh, Yungeun and Kim, Taewon and Choi, Seungman and Kim, Byunggi and Yoo, Hongki and Kim, Seung-Woo and Kim, Young-Jin},
  journal={PhotoniX},
  volume={6},
  number={1},
  pages={1--15},
  year={2025},
  publisher={Springer}
}

@article{zhang2015sub,
  title={Sub-Rayleigh-diffraction imaging via modulating classical light},
  author={Zhang, Erfeng and Lin, Huizu and Liu, Weitao and Li, Quan and Chen, Pingxing},
  journal={Optics Express},
  volume={23},
  number={26},
  pages={33506--33513},
  year={2015},
  publisher={Optical Society of America}
}

@article{dertinger2009fast,
  title={Fast, background-free, 3D super-resolution optical fluctuation imaging (SOFI)},
  author={Dertinger, Thomas and Colyer, Ryan and Iyer, Gopal and Weiss, Shimon and Enderlein, J{\"o}rg},
  journal={Proceedings of the National Academy of Sciences},
  volume={106},
  number={52},
  pages={22287--22292},
  year={2009},
  publisher={National Academy of Sciences}
}

@article{tsang2017subdiffraction,
  title={Subdiffraction incoherent optical imaging via spatial-mode demultiplexing},
  author={Tsang, Mankei},
  journal={New Journal of Physics},
  volume={19},
  number={2},
  pages={023054},
  year={2017},
  publisher={IOP Publishing}
}

@article{morizur2010programmable,
  title={Programmable unitary spatial mode manipulation},
  author={Morizur, Jean-Fran{\c{c}}ois and Nicholls, Lachlan and Jian, Pu and Armstrong, Seiji and Treps, Nicolas and Hage, Boris and Hsu, Magnus and Bowen, Warwick and Janousek, Jiri and Bachor, Hans-A},
  journal={Journal of the Optical Society of America A},
  volume={27},
  number={11},
  pages={2524--2531},
  year={2010},
  publisher={Optical Society of America}
}

@article{fontaine2019laguerre,
  title={Laguerre-Gaussian mode sorter},
  author={Fontaine, Nicolas K and Ryf, Roland and Chen, Haoshuo and Neilson, David T and Kim, Kwangwoong and Carpenter, Joel},
  journal={Nature communications},
  volume={10},
  number={1},
  pages={1865},
  year={2019},
  publisher={Nature Publishing Group UK London}
}

@article{leach2002measuring,
  title={Measuring the orbital angular momentum of a single photon},
  author={Leach, Jonathan and Padgett, Miles J and Barnett, Stephen M and Franke-Arnold, Sonja and Courtial, Johannes},
  journal={Physical review letters},
  volume={88},
  number={25},
  pages={257901},
  year={2002},
  publisher={APS}
}

@article{chang2026super,
  title={Super-resolution imaging of limited-size objects},
  author={Chang, Taeyong and Adamo, Giorgio and Zheludev, Nikolay I},
  journal={arXiv preprint arXiv:2602.00719},
  year={2026}
}

@article{bhusal2022smart,
  title={Smart quantum statistical imaging beyond the Abbe-Rayleigh criterion},
  author={Bhusal, Narayan and Hong, Mingyuan and Miller, Ashe and Quiroz-Ju{\'a}rez, Mario A and Le{\'o}n-Montiel, Roberto de J and You, Chenglong and Maga{\~n}a-Loaiza, Omar S},
  journal={npj Quantum Information},
  volume={8},
  number={1},
  pages={83},
  year={2022},
  publisher={Nature Publishing Group UK London}
}

@article{chen2024multi,
  title={Multi-plane light conversion (MPLC) LP mode multiplexer based on grayscale maskless lithography},
  author={Chen, Xueli and Yan, Xiaoyu and Fang, Juncheng and Lei, Ting and Pang, Fufei},
  journal={Optics Express},
  volume={32},
  number={23},
  pages={40305--40314},
  year={2024},
  publisher={Optica Publishing Group}
}

@article{fang2021performance,
  title={Performance optimization of multi-plane light conversion (MPLC) mode multiplexer by error tolerance analysis},
  author={Fang, Juncheng and Bu, Jing and Li, Jinpei and Lin, Chuxuan and Kong, Aru and Yin, Xiang and Luo, Haipeng and Song, Xiaolu and Xie, Zhenwei and Lei, Ting and others},
  journal={Optics Express},
  volume={29},
  number={23},
  pages={37852--37861},
  year={2021},
  publisher={Optical Society of America}
}

@article{zhou2019quantum,
  title={Quantum-limited estimation of the axial separation of two incoherent point sources},
  author={Zhou, Yiyu and Yang, Jing and Hassett, Jeremy D and Rafsanjani, Seyed Mohammad Hashemi and Mirhosseini, Mohammad and Vamivakas, A Nick and Jordan, Andrew N and Shi, Zhimin and Boyd, Robert W},
  journal={Optica},
  volume={6},
  number={5},
  pages={534--541},
  year={2019},
  publisher={Optical Society of America}
}

@article{menart2026definition,
  title={Definition of Multi-Parameter Super-Resolution for Optical Telescopes},
  author={Menart, Peter and Choi, Seungman and Jacob, Zubin},
  journal={arXiv preprint arXiv:2610.14286},
  year={2026}
}

@article{boucher2020spatial,
  title={Spatial optical mode demultiplexing as a practical tool for optimal transverse distance estimation},
  author={Boucher, Pauline and Fabre, Claude and Labroille, Guillaume and Treps, Nicolas},
  journal={Optica},
  volume={7},
  number={11},
  pages={1621--1626},
  year={2020},
  publisher={Optical Society of America}
}

@article{choi2026poisson,
  title={Poisson wavefront imaging in photon-starved scenarios},
  author={Choi, Seungman and Menart, Peter and Schramka, Andrew and Bauer, Leif and Aggarwal, Vaneet and Park, In-Yong and Jacob, Zubin},
  journal={Journal of the Optical Society of America B},
  volume={43},
  number={2},
  pages={231--238},
  year={2026},
  publisher={Optica Publishing Group}
}

@article{choi2025photon,
  title={Photon-starved imaging through turbulence at the diffraction limit},
  author={Choi, Seungman and Menart, Peter and Schramka, Andrew and Jape, Shubhankar and Bauer, Leif and Park, In-Yong and Jacob, Zubin},
  journal={arXiv preprint arXiv:2510.22806},
  year={2025}
}

@article{vrehavcek2017multiparameter,
  title={Multiparameter quantum metrology of incoherent point sources: Towards realistic superresolution},
  author={{\v{R}}eha{\v{c}}ek, J and Hradil, Z and Stoklasa, B and Pa{\'u}r, M and Grover, J and Krzic, A and S{\'a}nchez-Soto, LL},
  journal={Physical Review A},
  volume={96},
  number={6},
  pages={062107},
  year={2017},
  publisher={APS}
}

@article{santamaria2024single,
  title={Single-photon sub-Rayleigh precision measurements of a pair of incoherent sources of unequal intensity},
  author={Santamaria, Luigi and Sgobba, Fabrizio and Lupo, Cosmo},
  journal={Optica Quantum},
  volume={2},
  number={1},
  pages={46--56},
  year={2024},
  publisher={Optica Publishing Group}
}

@article{kingma2014adam,
  title={Adam: A method for stochastic optimization},
  author={Kingma, Diederik P and Ba, Jimmy},
  journal={arXiv preprint arXiv:1412.6980},
  year={2014}
}

@article{choi2025super,
  title={Super-Resolution Quantum Imaging by Adaptively Maximizing Single Photon Fisher Information},
  author={Choi, Hyunsoo and Jacob, Zubin and Baac, Hyoung Won},
  journal={ACS Photonics},
  volume={12},
  number={10},
  pages={5522--5527},
  year={2025},
  publisher={ACS Publications}
}

@article{deng_diffraction_2022,
	title = {Diffraction efficiency of {MEMS} phase light modulator, {TI}-{PLM}, for quasi-continuous and multi-point beam steering},
	volume = {13},
	number = {6},
	journal = {Micromachines},
	author = {Deng, Xianyue and Tang, Chin-I and Luo, Chuan and Takashima, Yuzuru},
	year = {2022},
	note = {Number: 6
Publisher: MDPI},
	pages = {966},
}

@article{chao_high-brightness_2023,
  title={High-brightness holographic projection},
  author={Chao, Brian and Gopakumar, Manu and Choi, Suyeon and Wetzstein, Gordon},
  journal={Optics Letters},
  volume={48},
  number={15},
  pages={4041--4044},
  year={2023},
  publisher={Optica Publishing Group}
}
\clearpage 
\appendix


\section*{Supplementary material}

# Adaptive Information Mining: Approaching the Quantum Limit of Super-Resolution Imaging [Supplementary Document]

**Contents**



## 1. ADAPTIVE INFORMATION MINING FORWARD MODEL

To simulate anisoplanatic phase modulation and image formation in the AIM framework, we employ a multi-slice wave propagation model based on the angular spectrum method (ASM). We consider a scene consisting of $N$ incoherent point sources $\mathbf{\Theta} = \{(x_n, y_n, b_n)\}_{n=1}^{N}$, where $(x_n, y_n)$ is the source location and $b_n$ is the brightness. The electric field for the $n$-th source at the pupil plane is defined as

$$E_n^{\text{pupil}}(u,v) = \sqrt{b_n} \cdot A(u,v) \cdot \exp[-j\frac{2\pi}{\lambda f}(x_n u + y_n v)] \tag{S1}$$

where $(u, v)$ are the spatial coordinates, $A(u, v)$ is the pupil aperture, $\lambda$ is the operating wavelength, and $f$ is the focal length of the system. The propagation of an arbitrary field $E_{\text{in}}(u, v)$ over an axial distance $z$ is modeled by the ASM operator $\mathcal{A}_z$ [1, 2]:

$$E_{\text{out}}(u,v) = \mathcal{A}_z[E_{\text{in}}(u,v)] = \mathcal{F}^{-1}[\mathcal{F}[E_{\text{in}(u,v)}] \cdot \mathcal{H}(k_x, k_y; z)] \tag{S2}$$

where, $\mathcal{F}$ and $\mathcal{F}^{-1}$ denote the 2D Fourier transform and its inverse, respectively. The transfer function is given by $\mathcal{H}(z) = \exp[j\frac{2\pi z}{\lambda}\sqrt{1-\lambda^2(k_x^2+k_y^2)}]$, where $(k_x, k_y)$ are the spatial frequency components corresponding to the spatial coordinates $(u, v)$. This function is defined within the region $k_x^2 + k_y^2 \leq 1/\lambda^2$ and set to zero otherwise, thereby excluding non-propagating evanescent components.

The field evolves sequentially through the optical train, undergoing free space propagation and phase modulation at each functional plane. The resulting field at the detector $E_n^{\text{det}}$ is determined by the full propagation operator $\mathcal{P}$:

$$E_n^{\text{det}}(\mathbf{\Phi};\mathbf{\Theta}) = \mathcal{P}[E_n^{\text{pupil}}(\mathbf{\Theta})] = \mathcal{A}_{z_4}[\mathcal{A}_{z_3}[\mathcal{A}_{z_2}[\mathcal{A}_{z_1}[E_n^{\text{pupil}}(\mathbf{\Theta})] \cdot L] \cdot \Gamma^{(1)}] \cdot \Gamma^{(2)}] \tag{S3}$$

where $L(u, v) = \exp[-j\frac{\pi}{\lambda f}(u^2 + v^2)]$ is the phase profile of the imaging lens, and $\mathbf{\Phi} = \{\Gamma^{(1)}, \Gamma^{(2)}\}$ represents the first and second SLM phase pattern, respectively. Here, $z_1, z_2, z_3$ and $z_4$ denote the axial distances between the sequential optical elements (pupil to lens, lens to SLM1, SLM1 to SLM2, and SLM2 to detector). The total mean photon-count vector $\boldsymbol{\mu}(\mathbf{\Phi}; \mathbf{\Theta})$ is obtained by the incoherent sum of intensities from all sources, $\sum_{n=1}^{N} |E_n^{\text{det}}(\mathbf{\Phi}; \mathbf{\Theta})|^2$.

## 2. MULTI POINT SOURCES MATCHING AND ERROR EVALUATION

To quantitatively assess the estimation accuracy for a multi-source scene, a formal correspondence between the ground truth set and the estimated set must be established. Given $N$ point sources, we

construct a cost matrix $C \in \mathbb{R}^{N \times N}$, where each element $C_{i,j}$ represents the dissimilarity between the $i$-th GT point and the $j$-th estimate. The cost function integrates both spatial Euclidean distance and brightness discrepancy:

$$C_{i,j} = \alpha \sqrt{(y_i - y_j)^2 + (x_i - x_j)^2} + \frac{\beta}{2} |b_i - b_j| \tag{S4}$$

In this formulation, $\alpha$ is a fixed spatial weight (set to 1.0). To balance the different physical scales of coordinates (pixels) and intensity (photon counts), the weight $\beta$ is dynamically scaled based on the global statistics of the scene:

$$\beta = \frac{\langle \Delta d_{xy} \rangle}{\langle \Delta b \rangle} = \frac{\sum_{i,j} \sqrt{(y_i - y_j)^2 + (x_i - x_j)^2}}{\sum_{i,j} |b_i - b_j|} \tag{S5}$$

The optimal one-to-one mapping is then determined by solving the linear sum assignment problem using the Hungarian algorithm. This combinatorial optimization identifies a binary assignment matrix $X$ that minimizes the total cost:

$$\min_{X} \sum_{i=1}^{N} \sum_{j=1}^{M} C_{i,j} X_{i,j} \tag{S6}$$

subject to the constraints $\sum_j X_{i,j} \leq 1$ and $\sum_i X_{i,j} \leq 1$, ensuring a unique one-to-one mapping where each GT source is paired with at most one estimate, and vice versa. This process ensures that the reported performance metrics, such as RMSE, are calculated from the most physically feasible associations.

## 3. OPTIMALITY CRITERION SELECTION

To clarify the role of optimality criteria in adaptive measurement design, we interpret the two most widely used criteria, A and D-optimality, through the Fisher information matrix (FIM) analysis. In particular, we show how different optimality choices correspond to distinct geometric contractions of uncertainty in the parameter space.

We consider a measurement model in which the measurement mean vector $\boldsymbol{\mu} \in \mathbb{R}^D$ depends on a parameter vector $\boldsymbol{\Theta} \in \mathbb{R}^P$, where $D$ denotes the number of detector pixels and $P$ denotes the number of parameters of interest.

The sensitivity of the measurements to the parameters is described by the Jacobian matrix:

$$\frac{\partial \boldsymbol{\mu}}{\partial \boldsymbol{\Theta}} \in \mathbb{R}^{D \times P}, \tag{S7}$$

whose $(d, p)$-th entry represents the partial derivative of the $d$-th measurement mean with respect to the $p$-th parameter.

Under a Poisson noise model, the FIM is given by

$$\mathbf{J} = \left( \frac{\partial \boldsymbol{\mu}}{\partial \boldsymbol{\Theta}} \right)^T \mathrm{diag}\left( \frac{1}{\boldsymbol{\mu}} \right) \left( \frac{\partial \boldsymbol{\mu}}{\partial \boldsymbol{\Theta}} \right) \in \mathbb{R}^{P \times P}, \tag{S8}$$

which quantifies the amount of information that the measurements provide about the unknown parameters.

We define the estimation error as

$$\delta \boldsymbol{\Theta} = \hat{\boldsymbol{\Theta}} - \boldsymbol{\Theta}_{\mathrm{GT}}, \tag{S9}$$

where $\hat{\boldsymbol{\Theta}}$ denotes the estimated parameter vector and $\boldsymbol{\Theta}_{\mathrm{GT}}$ denotes the ground truth parameters.

Under the standard Cramer-Rao lower bound (CRLB) interpretation, the estimation error is approximated as a zero-mean Gaussian random vector,

$$\delta \boldsymbol{\Theta} \sim \mathcal{N}(\mathbf{0}, \boldsymbol{\Sigma}), \tag{S10}$$

with covariance matrix $\boldsymbol{\Sigma}$.

When the covariance matrix is approximated by the inverse FIM, $\boldsymbol{\Sigma}^{-1} \approx \mathbf{J}$, the probability density function of the estimation error becomes

$$p(\delta \boldsymbol{\Theta}) \propto \exp\left( -\frac{1}{2} \delta \boldsymbol{\Theta}^T \mathbf{J} \delta \boldsymbol{\Theta} \right). \tag{S11}$$

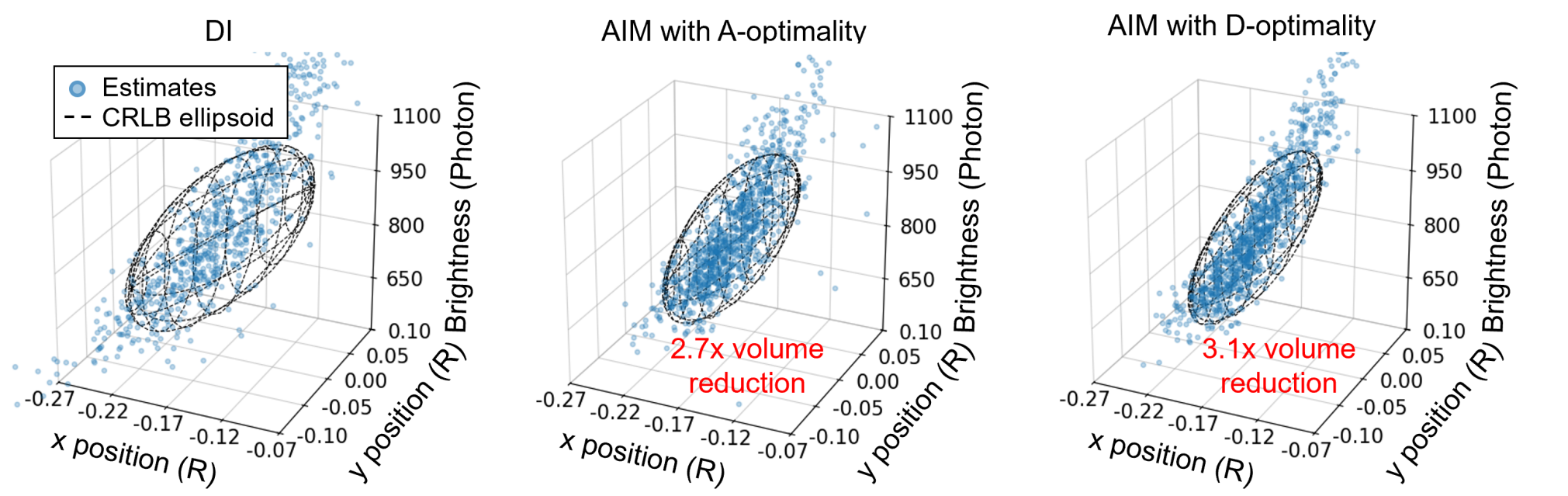


**Fig. S1.** A- and D-optimality comparison. Monte Carlo estimates (blue markers) and CRLB ellipsoids (dashed) illustrate estimation uncertainty under DI and AIM optimized with A- and D-optimality. D-optimality yields a larger reduction in CRLB ellipsoid volume. The Monte Carlo analysis is conducted using 1,000 trials for a two-point-source scene with separation $0.33R$ in $x$-axis, brightness ratio 16, and a total photon number of 14,000. Only the dim source is shown for clarity.

Iso-density contours of this distribution therefore satisfy

$$\delta\mathbf{\Theta}^T \mathbf{J} \delta\mathbf{\Theta} = c \tag{S12}$$

which defines an uncertainty (Confidence) ellipsoid in the parameter space. Without loss of generality, we set $c = 1$, which preserves the shape of the CLRB ellipsoid.

To interpret this uncertainty ellipsoid geometrically, we express the FIM via its eigen-decomposition,

$$\mathbf{J} = \mathbf{Q}\mathbf{\Lambda}\mathbf{Q}^T, \tag{S13}$$

where $\mathbf{Q} \in \mathbb{R}^{P\times P}$ contains the orthonormal eigenvectors and $\mathbf{\Lambda} = \mathrm{diag}(\lambda_1, \ldots, \lambda_P)$ contains the corresponding eigenvalues.

Under the coordinate transformation $\delta\mathbf{\Theta} = \mathbf{Q}\mathbf{z}$, the quadratic form becomes

$$\delta\mathbf{\Theta}^T \mathbf{J} \delta\mathbf{\Theta} = z^T \mathbf{\Lambda} z = \sum_{p=1}^{P} \lambda_p z_p^2 = 1, \tag{S14}$$

revealing that the principal axes of the ellipsoid are aligned with the eigen-directions of the FIM and that the axis lengths scale as $1/\sqrt{\lambda_p}$. Within this geometric interpretation, different optimality criteria correspond to different ways of contracting the uncertainty ellipsoid.

A-optimality seeks to minimize the trace of the inverse FIM,

$$\min \mathrm{Tr}(\mathbf{J}^{-1}) = \min \sum_{p=1}^{P} \frac{1}{\lambda_p}. \tag{S15}$$

Geometrically, this criterion minimizes the average variance across all parameter directions. As a result, A-optimality may mainly reduce uncertainty along directions that most efficiently decrease the trace, potentially leading to anisotropic contraction when the eigenvalues differ significantly.

In contrast, D-optimality maximizes the determinant of the FIM, or equivalently minimizes the determinant of its inverse,

$$\max \det(\mathbf{J}) \iff \min \prod_{p=1}^{P} \frac{1}{\lambda_p}. \tag{S16}$$

Since the determinant of the FIM is inversely proportional to the volume of the uncertainty ellipsoid, D-optimality explicitly minimizes the overall uncertainty volume, promoting a more balanced reduction across all parameters of interest.

Figure S1 illustrates these geometric differences by jointly visualizing Monte Carlo estimation results and the corresponding CRLB ellipsoids for a two-point-source scenario in the three-dimensional parameter space; $(x, y)$ position and brightness level.

The blue markers represent 1000 Monte Carlo estimates of the dim source parameters, while the dashed surfaces denote the CRLB ellipsoids predicted by the FIM. For direct imaging (DI),

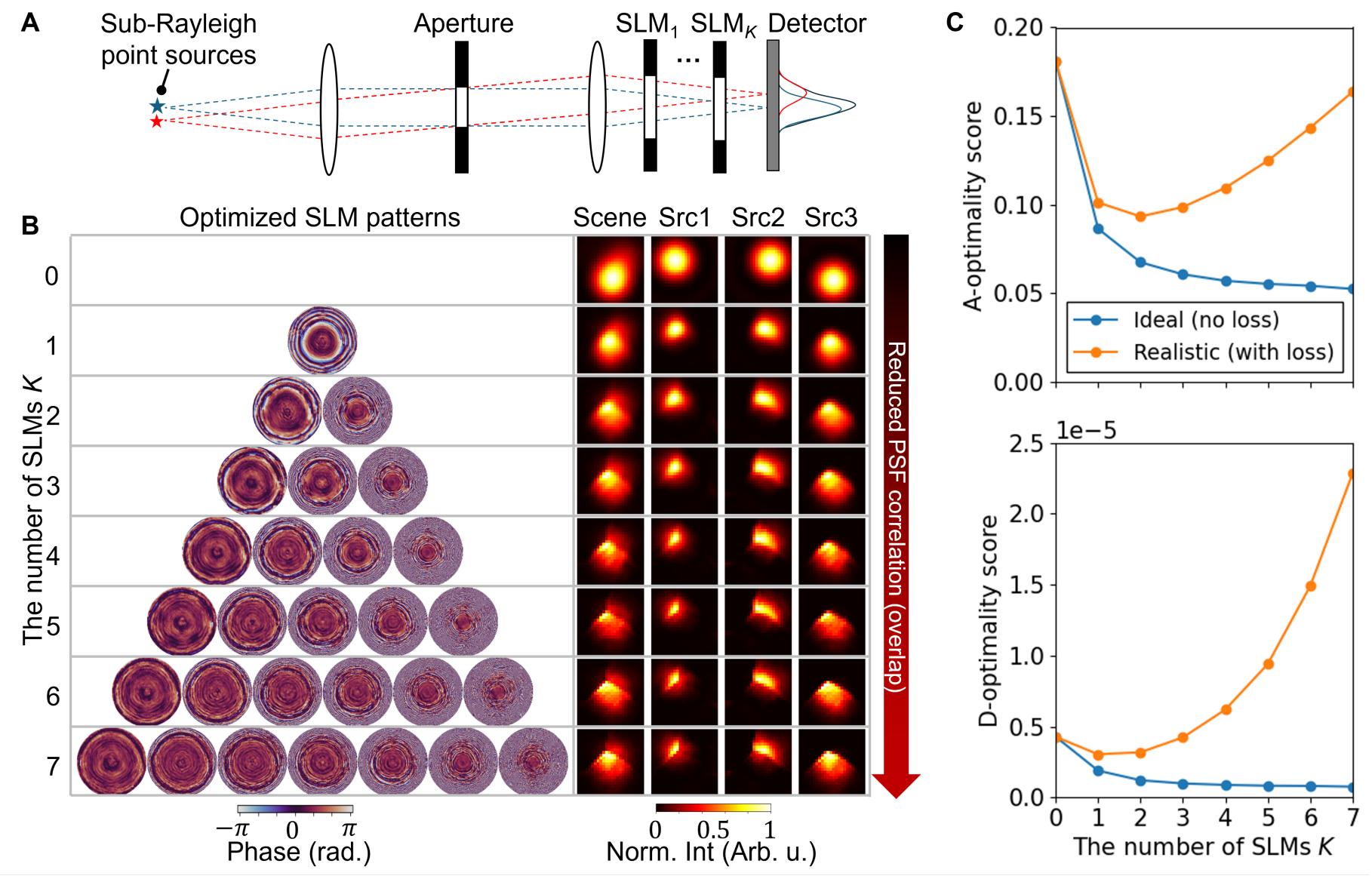


**Fig. S2.** Performance versus number of SLMs $K$ in the AIM setup. (A) System schematic illustrating sequential SLM modulation. (B) Optimized SLM phase patterns using D-optimality and corresponding scene responses and PSFs, showing reduced PSF correlation with increasing $K$. (C) A- and D-optimality scores versus $K$ for ideal (blue) and realistic (orange) systems with diffraction efficiency ($\sim 0.8$ per SLM). While increasing $K$ improves performance in the lossless case (lower A- and D-optimality scores), diffraction efficiency loss introduces a trade-off, yielding optimal performance at small $K$ (typically $K \approx 2$). Here, the SLM patterns are optimized for a scene consisting of three point sources with a uniform separation of $0.66R$ and brightness values $[1, 5, 15]$.

the estimation scatter closely follows the relatively large CRLB ellipsoid, indicating substantial uncertainty across position and brightness parameters.

Our Adaptive Information Mining (AIM) designs based on A and D-optimality significantly contract the CRLB ellipsoid and the associated estimation scatter, demonstrating improved estimation performance over DI. However, the D-optimal design yields a more pronounced reduction in the CRLB ellipsoid volume than the A-optimal design, consistent with its determinant-based objective. Since our goal is to achieve balanced uncertainty reduction across all parameters, including both source position and brightness, we adopt D-optimality as the primary design criterion for AIM in this work.

## 4. DESIGN TRADE-OFFS IN MULTI-SLMS ADAPTIVE MEASUREMENTS

Figure S2 illustrates how the number of sequential SLMs $K$ influences the performance of the AIM framework. As shown in Fig. S2(A), each additional SLM introduces new degrees of freedom for modulating the optical field prior to detection, enabling more flexible shaping at the detector plane. The optimized SLM phase patterns and corresponding scene responses in Fig. S2(B) demonstrate that increasing $K$ progressively reduces the correlation between the resulting per-source point spread functions (PSFs). This reduction in PSF correlation improves the distinguishability and localization of closely spaced point sources, leading to enhanced estimation performance in sub-Rayleigh imaging.

This trend is quantitatively captured in Fig. S2(C), which reports the A- and D-optimality scores as a function of $K$. In the ideal lossless case, increasing the number of SLMs $K$ consistently improves both optimality metrics, reflecting the benefit of increased modulation freedom. However, the marginal performance gain diminishes as $K$ grows, with the largest improvement occurring when introducing the first and second SLM.

In realistic systems, where each SLM has diffraction efficiency loss (0.8 per SLM in this simulation), the trade-off between information gain and photon loss becomes significant. Under this constraint, increasing $K$ beyond a small number no longer yields performance improvement, and

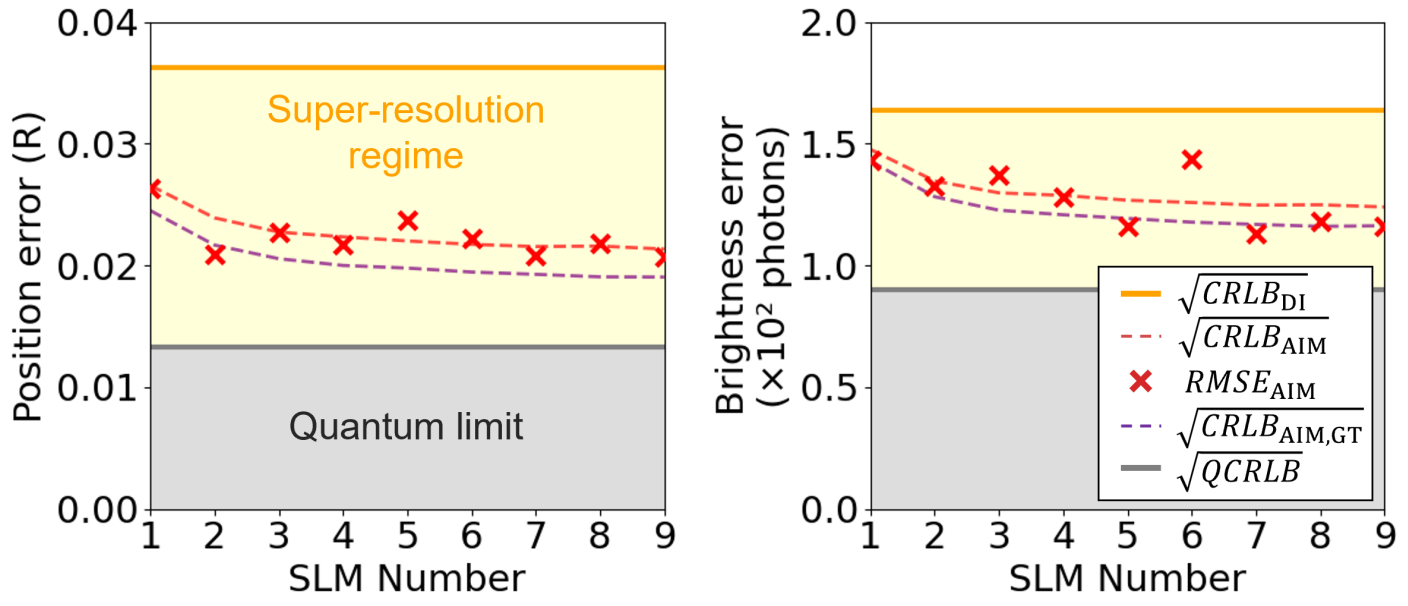


**Fig. S3.** Super-resolution performance of AIM, with position and brightness RMSE plotted versus the number of SLM patterns (red crosses). The scene consists of three point sources with brightness ratios of [1, 5, 20]. AIM results are obtained using five sequential measurements at a source separation of 0.5 R, with a total photon budget of 14,000. Errors are averaged over three sources across 50 Monte Carlo trials per condition. The yellow and gray lines denote the classical and quantum CRLB, respectively. The red and purple dashed curves represent the AIM CRLB under adaptive and non-adaptive measurements (with known ground truth), a benchmark highlighting the potential for the AIM RMSE to approach as the number of measurements increases.

both A- and D-optimality scores are optimal at $K \approx 2$. These results provide practical guidance for selecting the number of SLMs during the measurement design stage, where system-level considerations such as diffraction efficiency may be taken into account, while the main text focuses on loss-free performance.

Figure S3 evaluates the practical performance of AIM under ideal conditions. As discussed for the lossless case, increasing the number of SLM patterns provides additional degrees of freedom to modulate the complex field, allowing the estimation error to approach the quantum limit. Note that in this work, the FIM is optimized in a sequential, one-step (greedy) manner based only on the next measurement.

## 5. PERFORMANCE TRENDS ACROSS KEY SCENE PARAMETERS

Figure S4 summarizes how point-source estimation performance varies with key scene parameters. We consider a two point-source scene separated along the $x$-axis and evaluate estimation accuracy for each source's $(y, x)$ position and brightness. Three parameters are scanned separately: the inter-source separation, the total photon number, and the relative brightness ratio. For each scan, the performance of diffraction-limited DI is compared with that of the proposed AIM framework.

In Fig. S4, the left $2 \times 3$ panels show Monte Carlo estimation results (750 trials over the full scan range) for the $y$- and $x$-centroid positions and brightness, overlaid with the ground-truth values (solid lines). The right $2 \times 3$ panels report the corresponding root mean square errors (RMSE), together with the square root of CRLB, separately for the dim and bright sources.

We first examine the separation scan (Fig. S4A). As the separation decreases, estimation becomes progressively more challenging for both DI and AIM: the scatter around the ground truth increases, reflecting the growing correlation between the two sources. This behavior is inherent to sub-Rayleigh configurations, where overlapping PSFs reduce parameter identifiability. Despite this trend, AIM consistently exhibits a significantly lower variance than DI, with estimates more tightly clustered around the ground truth. This improvement is quantified in the error plots, where AIM achieves a clear reduction in RMSE across all parameters.

Importantly, AIM not only outperforms DI but also approaches its own CRLB benchmark. While DI shows noticeable discrepancies between RMSE and its CRLB in the sub-Rayleigh regimes, AIM's RMSE closely follows its theoretical CRLB across nearly all cases. This agreement suggests that the accumulated SLM-induced phase diversity in AIM effectively mitigates local minima, enabling convergence toward the global optimum assumed in the CRLB derivation.

The same qualitative trends are observed in the photon-number and brightness-ratio scans (Figs. S4B and S4C). As photon count decreases or brightness contrast becomes larger, estimation errors for the dim source increases for both methods; however, AIM consistently maintains lower RMSE and closer agreement with its CRLB than DI. Across all scanned parameters, AIM delivers

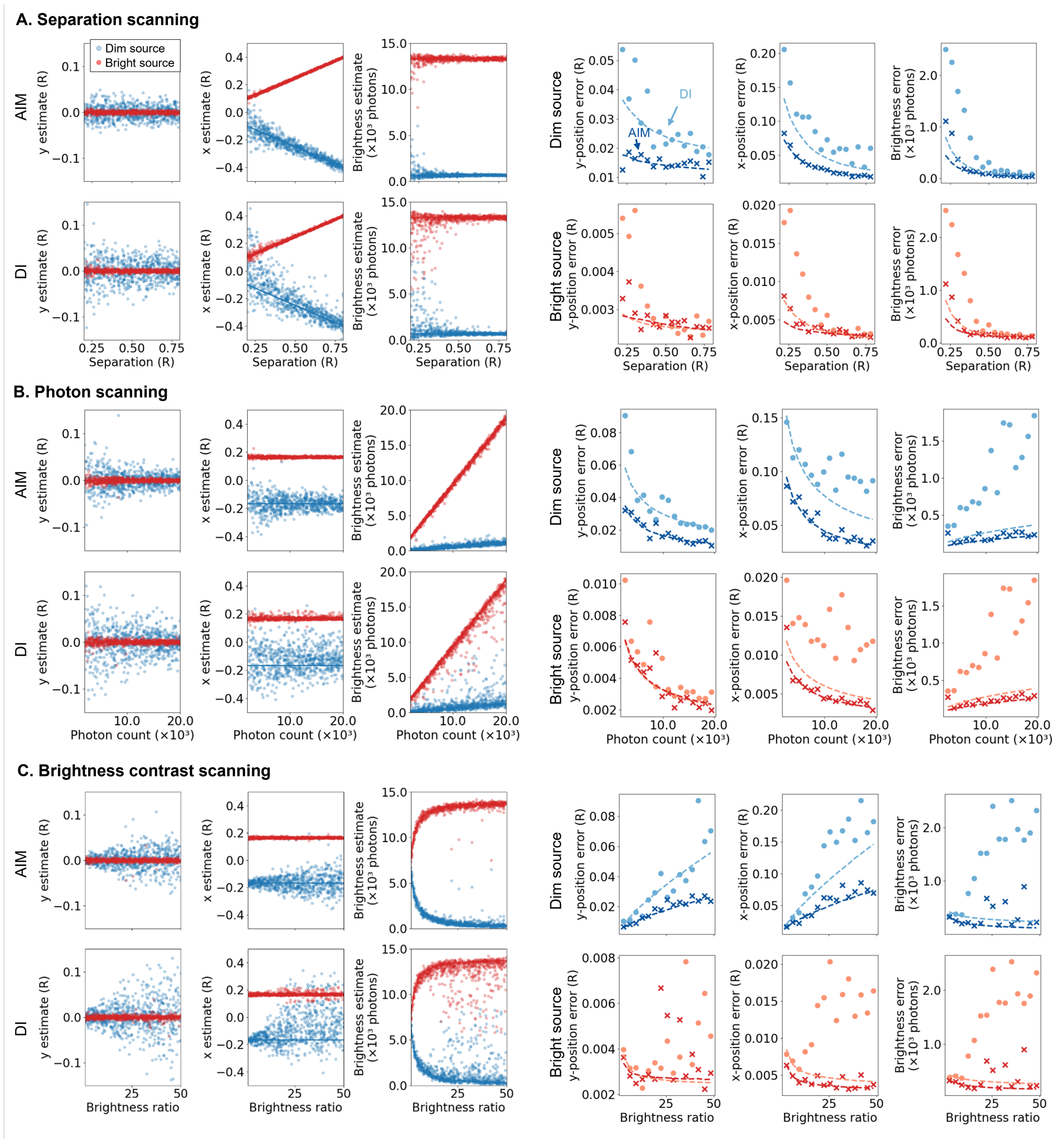


**Fig. S4.** Performance trends in two point-source parameter estimation. All panels consider a two-point-source scene separated along the $x$-axis. The left $2 \times 3$ panels show Monte Carlo estimates of $(y, x)$ position and brightness (scatter) overlaid with the ground truth (solid lines), while the right $2 \times 3$ panels report the corresponding RMSE together with CRLB predictions for the dim and bright sources. Panels (A)–(C) correspond to different parameter scans: (A) Separation scan: separation =0.2–0.8$R$ with 14,000 photons and a brightness ratio of [1 : 20]; (B) Photon-number scan: photon count =2,000–20,000 at separation =0.33$R$ with a brightness ratio of [1 : 16]; (C) Brightness-ratio scan: brightness ratio 1–50 at separation =0.33$R$ with 14,000 photons. A total of 750 Monte Carlo trials are used. For the error plots, results are grouped into 15 sub-ranges along the scanned parameter, with each data point representing approximately 50 trials.

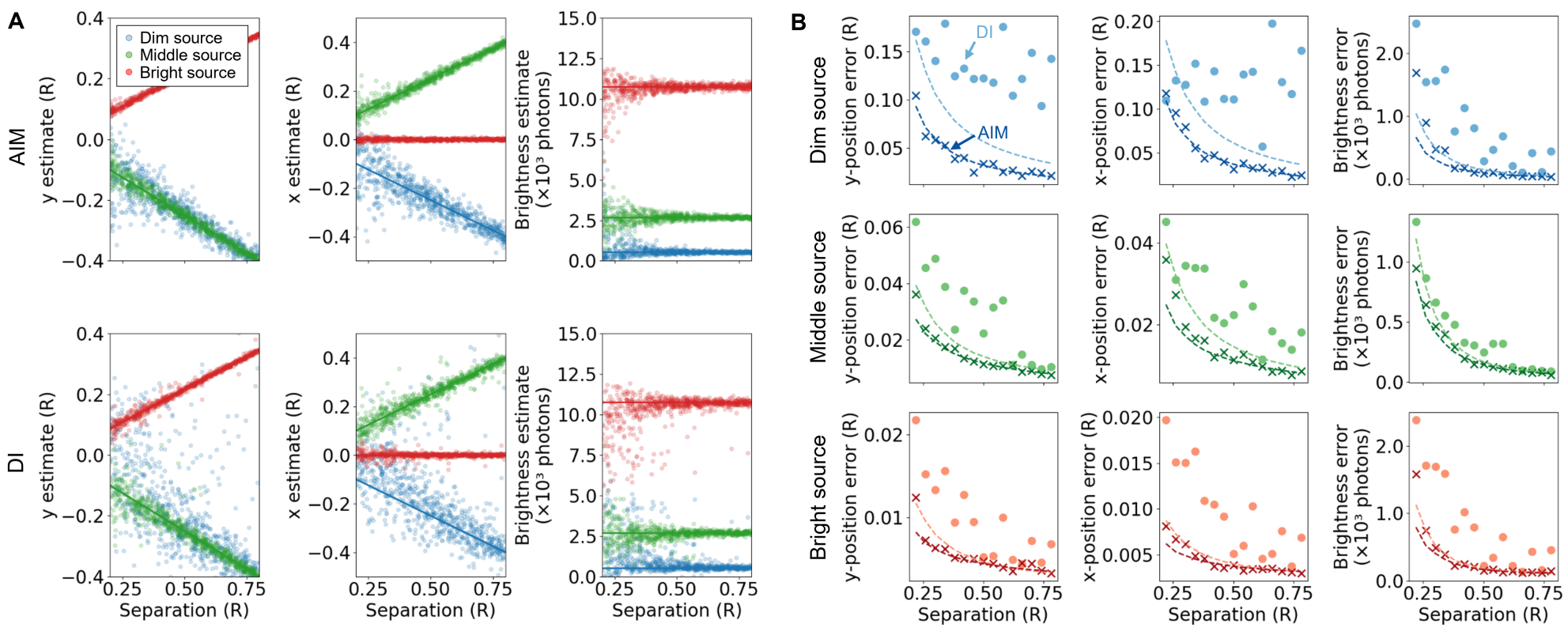


**Fig. S5.** Separation-scan comparison between AIM and DI for a three-point source scene. The source separation is varied from $0.2R$ to $0.8R$, with a total photon budget of 14,000 and a brightness ratio of $[1 : 5 : 20]$. (A) Monte Carlo estimates of the $y$- and $x$-positions and brightness (scatter points), overlaid with the ground truth values (solid lines). (B) Corresponding RMSE as a function of separation, together with the theoretical error bounds given by the CRLB. Darker dashed curves indicate the AIM CRLBs, while lighter dashed curves indicate the DI CRLBs. Scatter points indicate RMSE values grouped into 15 sub-ranges along the scanned separation axis, with each point representing approximately 50 Monte Carlo trials. A total of 750 Monte Carlo trials are used.

more robust, near CRLB-limited estimation for both centroid position and brightness, particularly in regimes where DI is most susceptible to source correlation and noise.

## 6. GENERALIZED RANDOM SCENE PERFORMANCE

While the above section focuses on the two-point-source scenario for clarity, the proposed AIM framework naturally extends to scenes with a larger number of sources. To demonstrate this generality, we evaluate AIM performance on a three-point-source scene with uniformly spaced separations, and compare it against conventional diffraction-limited DI.

Figure S5 presents a separation-scan study, where the source separation is varied from $0.20R$ to $0.80R$. Fig. S5A shows Monte Carlo estimates of source positions and brightness for AIM (top row) and DI (bottom row). Compared to DI, AIM estimates exhibit significantly reduced variance and remain more tightly aligned with the ground truth across all separations.

This improvement is quantified in S5B, which reports the corresponding RMSE for position and brightness estimation. The CRLB under AIM (darker dashed curves) is consistently lower than that of DI across all sources and parameters. Moreover, the empirical RMSE closely follows the CRLB. We confirmed that AIM enables robust and near-optimal estimation not only for two-point-source scenes, but also for generalized multi-source configurations.

Figure. S6 further extends this analysis to a more general scenario in which multiple point sources are randomly distributed in the scene. This case study illustrates how the AIM measurement configuration adapts to complex scenes. Across all tested realizations, the AIM measurements reshape the effective scene response such that mutual correlations between neighboring point sources are substantially reduced. This correlation-suppressing behavior directly translates into improved sub-Rayleigh imaging performance, demonstrating that AIM remains effective not only for structured multi-source configurations, but also for fully randomized scenes.

## 7. COMPARISON WITH QUANTUM LIMIT

We compare the estimation precision of DI and AIM against the theoretical Quantum Cramer–Rao Lower Bound (QCRLB) (see Appendix for derivation). Simulations were performed using a 3-point source scene, evaluating performance both as a function of source separation ($0.3R$–$1.5R$; Fig. SS7) and incoming photon count (2,000–20,000 with fixed separation $0.4R$; Fig. SS8).

To ensure stable convergence, both DI and AIM estimators were initialized using peak signal locations obtained from Richardson–Lucy deconvolution [3]. This initialization enables the

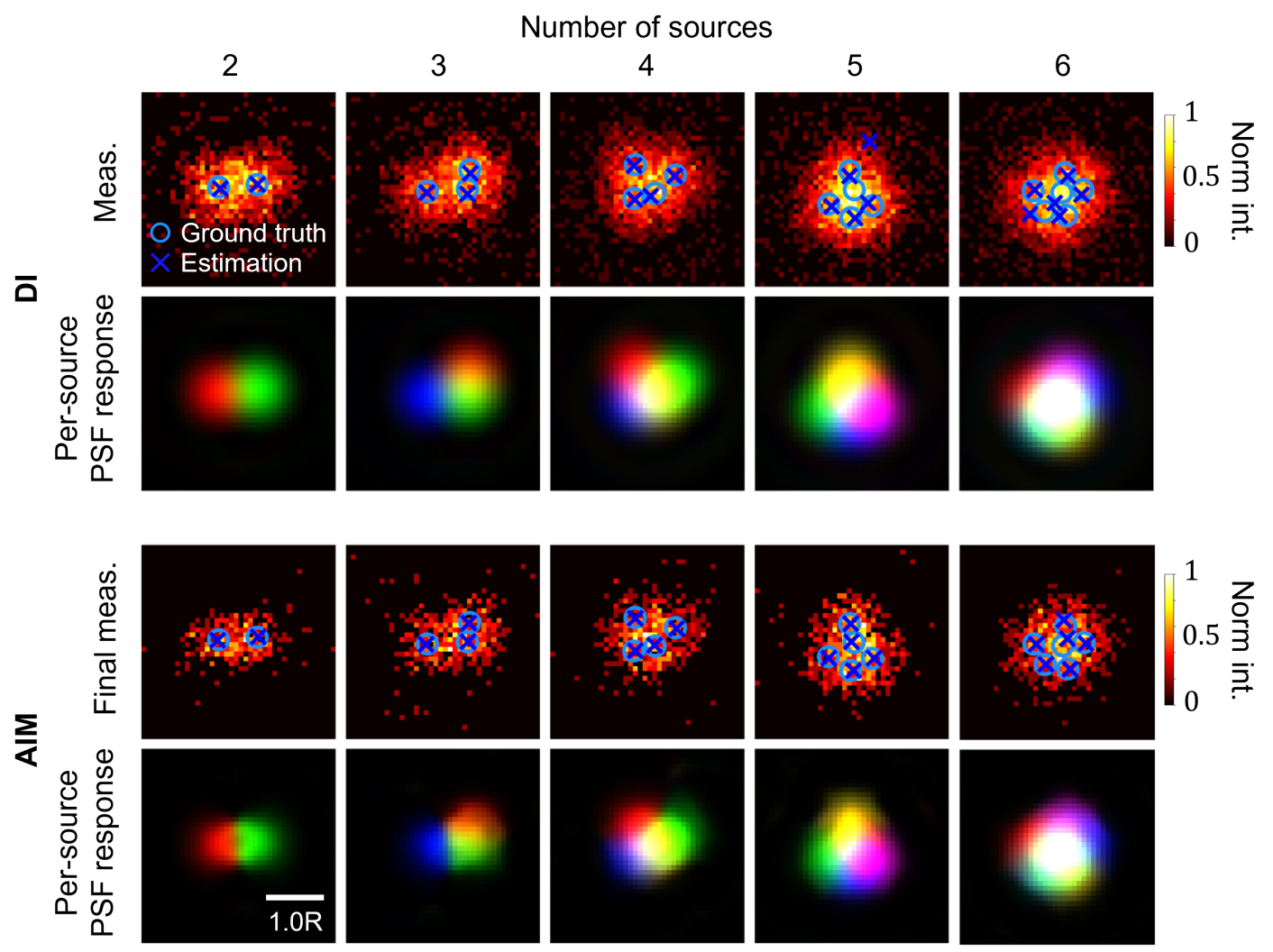


**Fig. S6.** Measurement images and source-resolved PSF responses for DI and AIM with randomly placed point sources. Color hue denotes source identity. Each source emits a total of 800 photons, which are then divided into five sequential measurements by the AIM framework. Sources are placed with a separation of < 1.0 R. Compared to DI, AIM reduces overlap between per-source PSF responses, improving source separability in multi-source scenes.

DI estimator to achieve improved convergence and follow the classical CRLB more closely, demonstrating that the classical limit can be approached through careful algorithmic design. In contrast, the AIM framework consistently tracks its theoretical CRLB regardless of initialization (see also Fig. SS5), reflecting the improved global convergence enabled by the information diversity of sequential measurements.

As shown in Fig. SS7A, AIM outperforms DI across all individual point-source parameters, substantially narrowing the gap to the quantum limit. This improvement is further consolidated in the ensemble average over the three sources (Fig. SS7B), where both DI and AIM closely follow their respective CRLBs, while the AIM bound remains consistently lower and approaches the QCRLB.

To quantify this improvement relative to the ultimate physical limit, we define an error reduction ratio using the quantum bound $\sqrt{QCRLB}$ as the baseline: $(\sqrt{CRLB_{\mathrm{DI}}} - \sqrt{QCRLB})/(\sqrt{CRLB_{\mathrm{AIM}}} - \sqrt{QCRLB})$. As summarized in Fig. SS7C, AIM reduces the excess estimation error relative to the quantum limit by a factor of approximately 2–3 for both position and brightness across the separation regime. Consistent behavior is also observed in the photon-scaling analysis of Fig. SS8.

## 8. ROBUSTNESS TO DETECTOR READ NOISE

AIM preserves its performance advantage over DI in the presence of detector read noise by adapting the underlying statistical model and FIM to account for both shot noise and read noise.

When both noise sources are present, the measurement likelihood is described by a Poisson–Gaussian mixture distribution,

$$p(y \mid \mu) = \sum_{k=0}^{\infty} \mathrm{Poisson}(k;\mu)\, \mathcal{N}(y-k;\, 0, \sigma^2), \tag{S17}$$

which is obtained by marginalizing over all possible photon counts $k$. While this model accurately captures the measurement statistics, it is computationally expensive to evaluate and analytically intractable for direct FIM calculation, as it requires summation over all discrete photon realizations.

A common and computationally efficient approximation is to replace the Poisson–Gaussian mixture model, whose mean is given by the expected signal intensity $\mu$ and whose variance is given by the combined shot noise and read noise contributions, $\mu + \sigma^2$. Under this approximation,

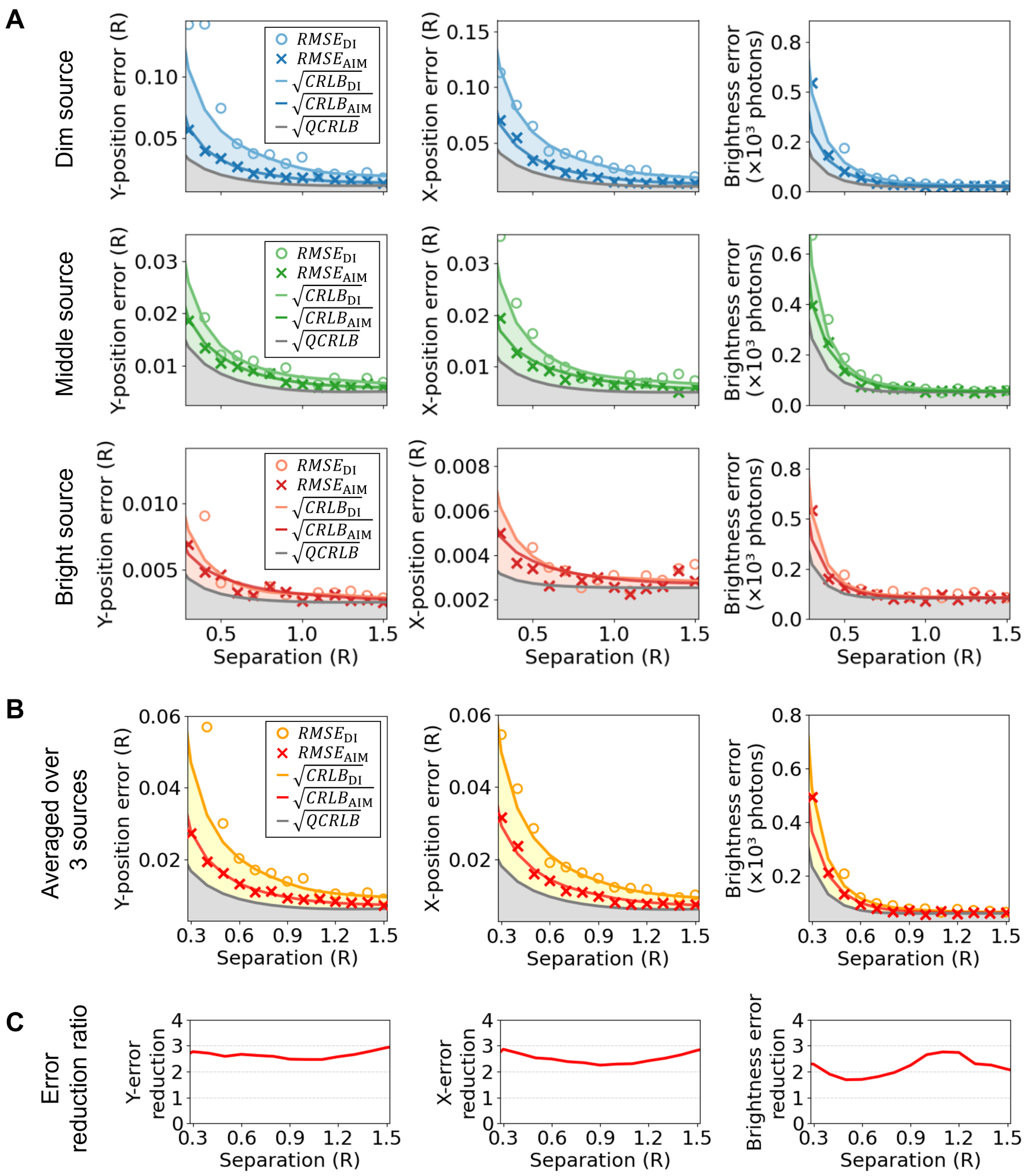

**Fig. S7.** Comparison of estimation performance for DI, AIM, and the Quantum CRLB in a 3-point source scene as a function of source separation. (A) Estimation accuracy for each individual point-source parameter and (B) the ensemble average over all three sources. (C) Error reduction ratio relative to the fundamental quantum limit, defined as $(\sqrt{CRLB_{\mathrm{DI}}} - \sqrt{QCRLB})/(\sqrt{CRLB_{\mathrm{AIM}}} - \sqrt{QCRLB})$, illustrating how AIM approaches the quantum-limited regime compared to DI. Simulations were performed over a separation range of 0.3–1.5$R$ with 50 trials per scatter point. The total photon count was fixed at 14,000 with brightness ratios of 1, 5, and 20. AIM used 5 adaptive measurements. For numerical stability, both DI and AIM estimators were initialized using peak locations obtained from Richardson–Lucy deconvolution [3].

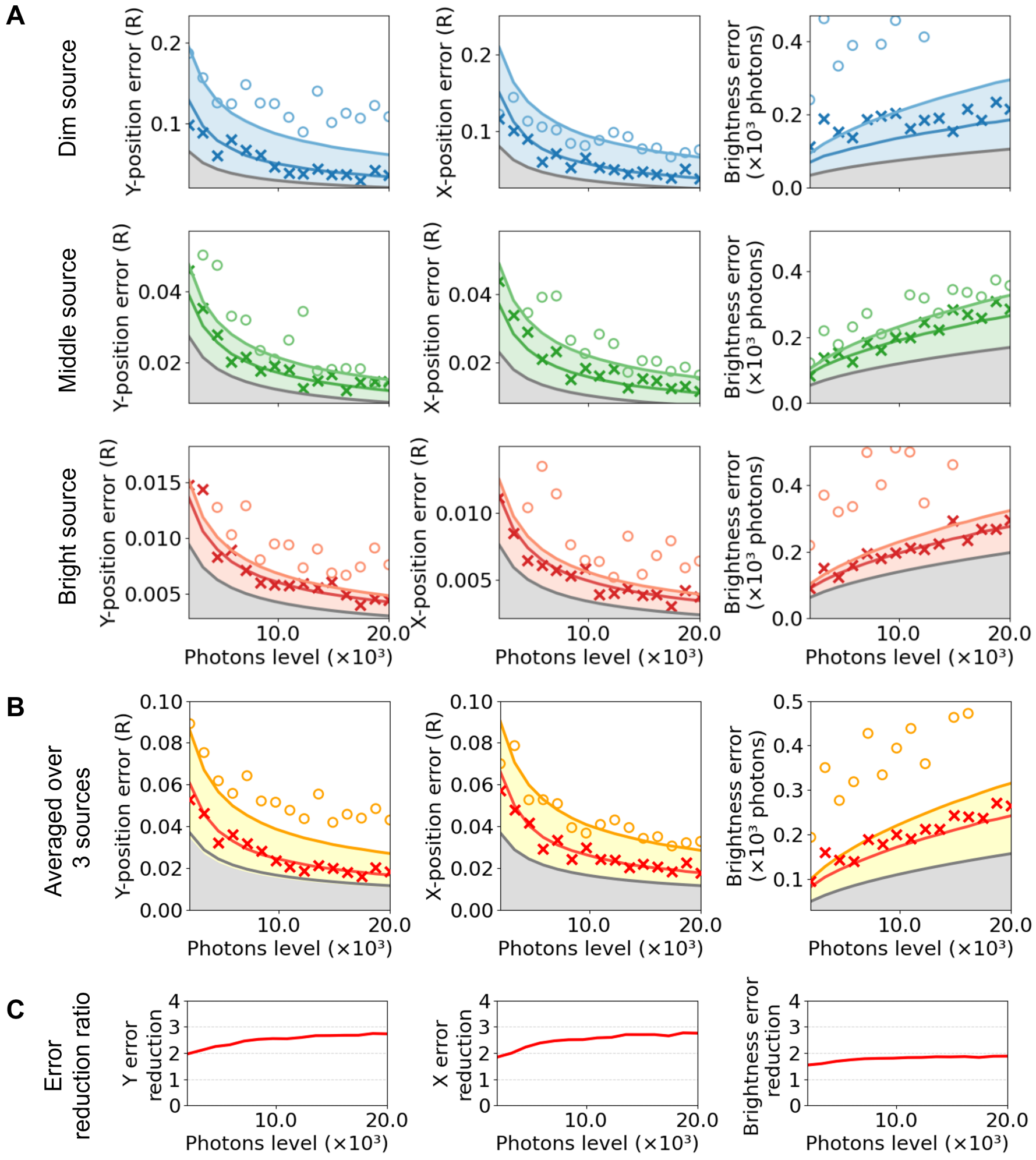


**Fig. S8.** Comparison of DI, AIM, and the Quantum CRLB in a 3-point source scene as a function of the incoming photon count. The separation was fixed at $0.4R$, and estimation performance is evaluated over a photon range of 2,000-20,000 under brightness ratios of 1, 5, and 20. Each scatter point corresponds to 50 Monte Carlo trials. AIM employs 5 adaptive measurements. For numerical stability, both DI and AIM estimators were initialized using peak locations obtained from Richardson–Lucy deconvolution [3].

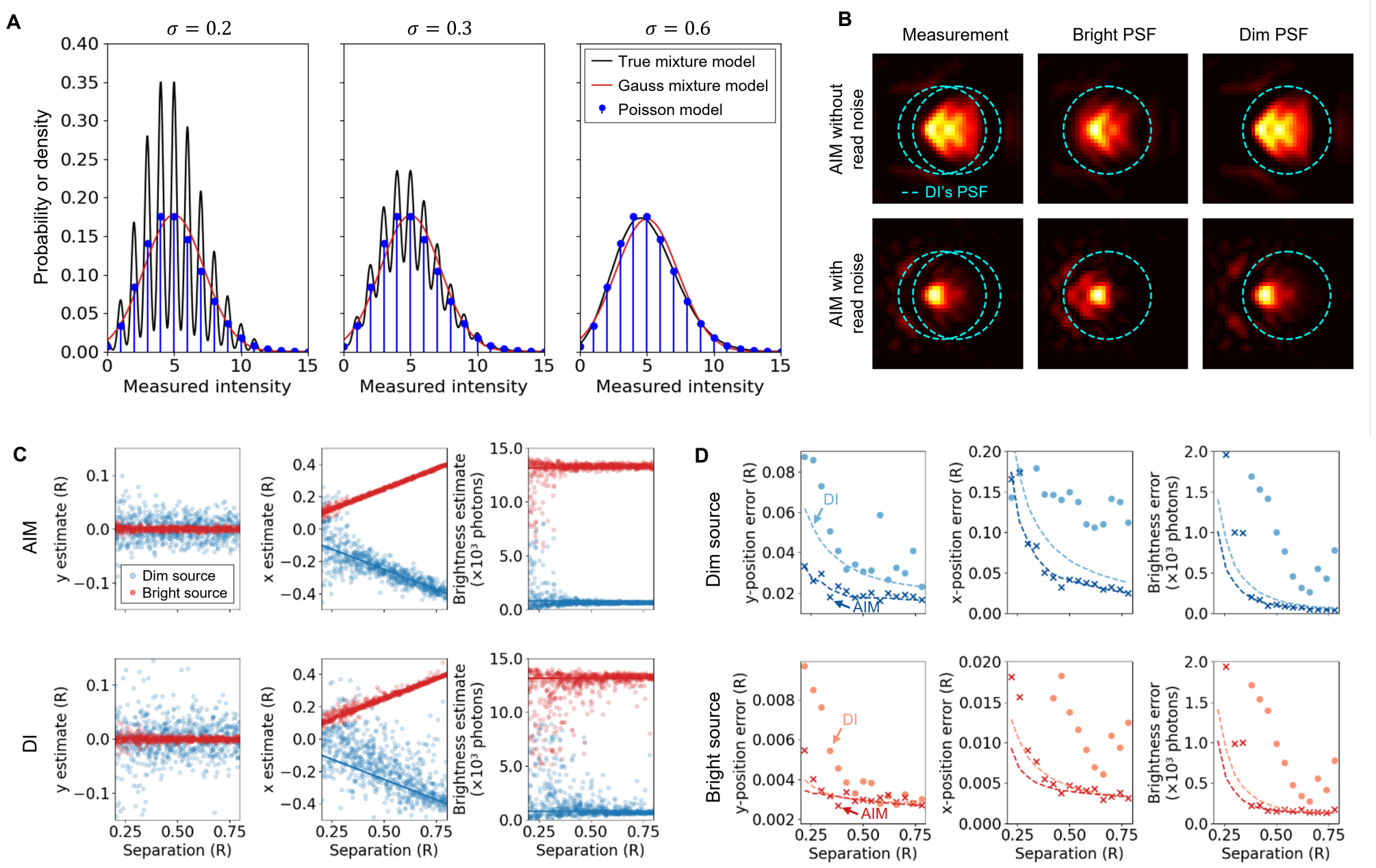


**Fig. S9.** Robustness to detector read noise in two-point-source parameter estimation. (A) Measured-intensity distributions for $\mu = 5$ under different read noise levels ($\sigma = 0.2, 0.3, 0.6$), comparing the true Poisson-Gaussian mixture model, a Gaussian approximation, and a Poisson model. (B) Representative measurements and corresponding PSFs for AIM without read noise, and AIM with read noise. In the presence of read noise, AIM adjusts the effective defocus to concentrate the PSF in smaller region to mitigate read noise effect. (C) Monte Carlo estimates of centroid positions and brightness as a function of source separation for AIM (top) and DI (bottom), overlaid with the ground truth. (D) Corresponding RMSE for position and brightness estimation (scatter) together with the CRLB (dashed). Results in (C) and (D) are shown for a separations from $0.2R$ to $0.8R$, with a total photon number of 14,000, a brightness ratio of $[1:20]$, and detector read noise with $\sigma = 0.6\mathrm{e}^-$.

the measurement probability density is expressed as

$$p(y \mid \mu) \approx \mathcal{N}\left(y;\, \mu,\ \mu + \sigma^2\right). \tag{S18}$$

Using this likelihood, the FIM with respect to the parameter vector $\mathbf{\Theta}$ can be approximated as

$$\mathcal{J}(\mathbf{\Theta}) \;\approx\; \left(\frac{\partial \boldsymbol{\mu}}{\partial \mathbf{\Theta}}\right)^{\top} \mathrm{diag}\left(\frac{1}{\boldsymbol{\mu} + \sigma^2}\right) \left(\frac{\partial \boldsymbol{\mu}}{\partial \mathbf{\Theta}}\right), \tag{S19}$$

where $\boldsymbol{\mu} \in \mathbb{R}^D$ is the expected photon count vector over $D$ detector pixels and $\mathbf{\Theta} \in \mathbb{R}^P$ is the vector of unknown scene parameters, as in Eq. S8.

Figure S9(A) compares the true Poisson–Gaussian mixture model with the Gaussian and Poisson approximations for $\mu = 5$ and different read noise levels. When the read noise is small, the true distribution exhibits sharp peaks centered at discrete photon numbers, closely resembling the Poisson model. As the read noise increases, these discrete features are smoothed out, and the Gaussian approximation more accurately captures the overall distribution envelope. This behavior indicates that the appropriate statistical model depends on the relative magnitude of the detector read noise.

Figure S9(B) shows representative measurement images and corresponding PSFs optimized using the FIM under noise-free and read-noise conditions. In the presence of read noise, AIM produces a more compact PSF compared to the noise-free case. This behavior arises from the two SLM phase patterns introducing a counter-defocus that effectively increases the system magnification, concentrating the signal into fewer detector pixels and reducing the relative impact of read noise. At the same time, excessive magnification reduces the information available for resolving closely spaced sources. AIM automatically balances this trade-off between signal concentration and spatial information by maximizing the overall Fisher information.

Figures S9(C) and S9(D) present Monte Carlo estimation results and the corresponding RMSE and CRLB under read-noise conditions, respectively. These results confirm that AIM consistently achieves lower estimation error than DI across a range of source separations, even in realistic case with detector read noise.

## 9. DETECTION PERFORMANCE

In our formulation, the parameter set of interest is $\mathbf{\Theta} = \{(x_n, y_n, b_n)\}_{n=1}^{N} \in \mathbb{R}^P$, where $x_n$, $y_n$, and $b_n$ denote the position and brightness of the $n$-th point source, and $P = 3N$ is the total number of unknown parameters. Importantly, the number of point sources $N$ is itself unknown in a realistic scenario and must be estimated from the measurements. Since $N$ is a discrete variable, it cannot be optimized directly using optimization scheme. Instead, during the estimation step, we consider multiple candidate models with different values of $N$. For each candidate, we perform parameter estimation and evaluate the corresponding negative log-likelihood $\mathcal{L}$. Because increasing $N$ generally improves data fitting by introducing additional degrees of freedom, likelihood alone is insufficient for selecting the correct model order. To balance model complexity and data fidelity, we adopt the Bayesian Information Criterion (BIC),

$$\mathrm{BIC}(P) = P \ln d + 2\mathcal{L} \tag{S20}$$

where $d$ denotes the number of effective data points. In our imaging model, $d$ is chosen as the number of detector pixels within the region of interest ($40 \times 40$ in this simulation). The candidate model with the minimum BIC is selected as the estimated number of point sources. The corresponding scene parameters are then used to optimize the SLM patterns in the subsequent AIM iteration.

Figure S10 compares the source-number detection performance of AIM and DI for a three-point-source scene, where the candidate source count is scanned from $N = 1$ to 5. Two scanning scenarios are considered: detection probability evaluated as a function of source separation and as a function of photon count. In both cases, AIM consistently achieves higher detection probability than DI particularly in challenging regimes where DI struggles to correctly infer the source number.

Although AIM does not directly optimize information with respect to the discrete source count $N$, it designs measurements that maximize information for each candidate scene configuration. By shaping the measurement to be highly sensitive to the parameters of individual point sources, AIM indirectly increases the information available for source presence and thereby enhances

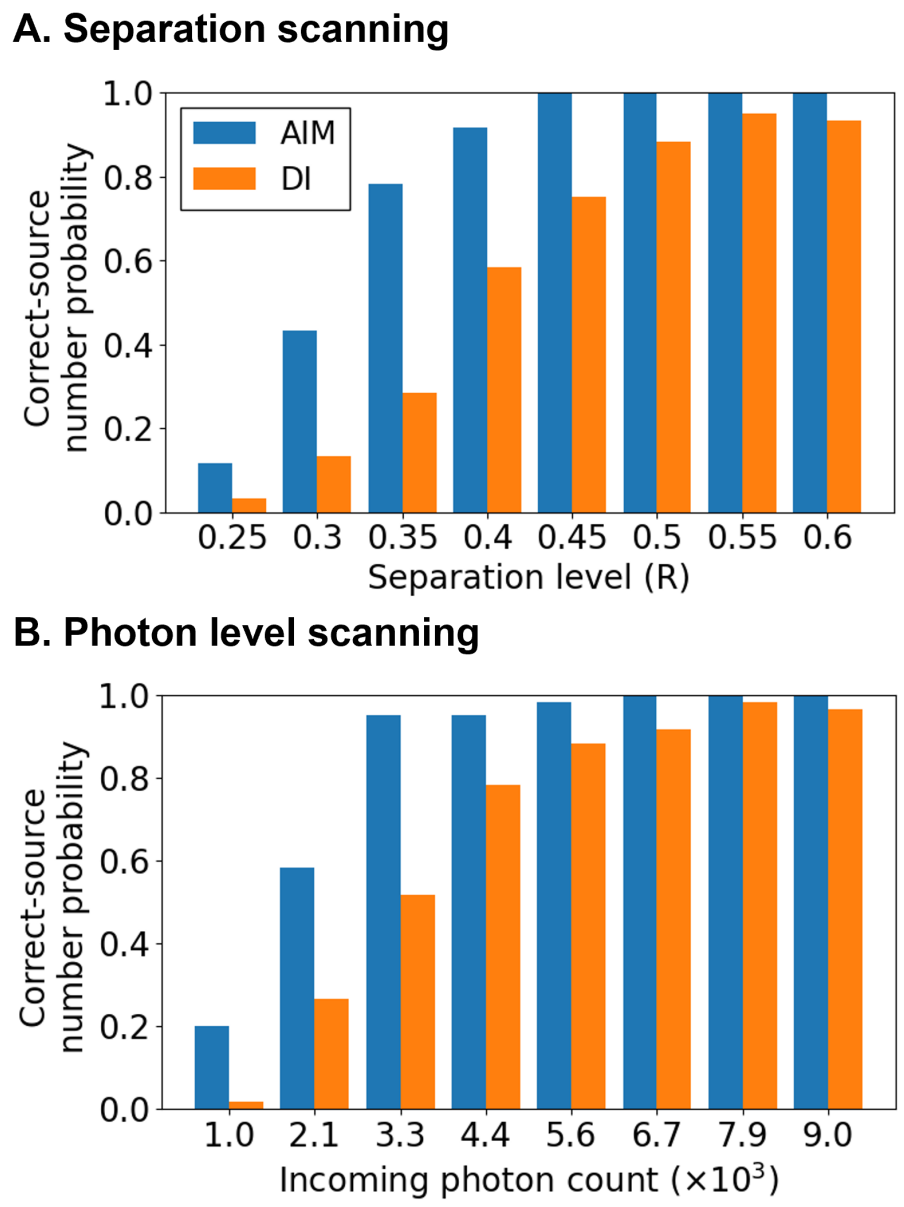


**Fig. S10.** Detection performance comparison between AIM and DI. (A) Probability of correctly estimating the source number versus separation (0.25–0.6 R) at 7000 photons. (B) Probability versus photon count (1000–9000 photons) at separation 0.5 R. Results correspond to a three-point-source scene with brightness ratios [1, 5, 20]. The source number is inferred by model selection over candidate values from 1 to 5. Each bar represents the average over 60 independent trials.

source detection. Consequently, incorrect source-count hypotheses are more effectively penalized during the BIC-based model selection process. These results demonstrate that information-driven measurement design not only improves localization accuracy, but also enhances the detectability of point sources under challenging conditions.

## 10. JOINT ESTIMATION OF SOURCE PARAMETERS AND TURBULENCE

In practical astronomical imaging, performance is fundamentally limited by atmospheric turbulence (seeing limit). To mitigate this, we extend the AIM framework to perform joint estimation of the scene parameters and the unknown turbulent wavefront [1, 2]. The augmented parameter set is defined as $\mathbf{\Theta}_{\text{total}} = \{\mathbf{\Theta}, \Phi_{\text{turb}}\}$, where $\mathbf{\Theta} = \{(x_n, y_n, b_n)\}_{n=1}^{N}$ represents the spatial and brightness parameters of the point sources, and $\Phi_{\text{turb}}$ denotes the phase of the turbulence. Crucially, while the estimation step accounts for the turbulence, the SLM optimization process remains dedicated to maximizing the Fisher information specifically with respect to the scene parameters $\mathbf{\Theta}$. Here, we assume that the AIM operation cycle is sufficiently faster than the atmospheric coherence time (time constant). This allows us to treat the turbulence as a "frozen" or static phase screen during the estimation and optimization process.

Fig. S11 compares two distinct optical configurations for this task. In Configuration 1 (Fig. S11A), the first SLM ($\Gamma^{(1)}$) is positioned at the aperture plane, while the second SLM ($\Gamma^{(2)}$) is placed downstream. In Configuration 2 (Fig. S11B), both SLMs are positioned downstream. While both architectures successfully recover the scene and turbulence parameters, they exhibit fundamentally different adaptive behaviors.

In Configuration 1, a remarkable "emergent compensation" effect is observed. Although the optimization objective is purely defined by information gain for the point sources, the system automatically allocates the degrees of freedom in the aperture-plane SLM to counteract the atmospheric turbulence. As shown in the residual phase analysis (Fig. S11C), the difference between $\Gamma^{(1)}$ and the ground-truth turbulence $\Phi_{\text{turb}}^{\text{GT}}$ reduces to a simple global tip/tilt and a small defocus component. This suggests that to maximize sensitivity for sub-Rayleigh localization, the AIM framework first restores the system to a diffraction-limited state by clearing isoplanatic aberra-

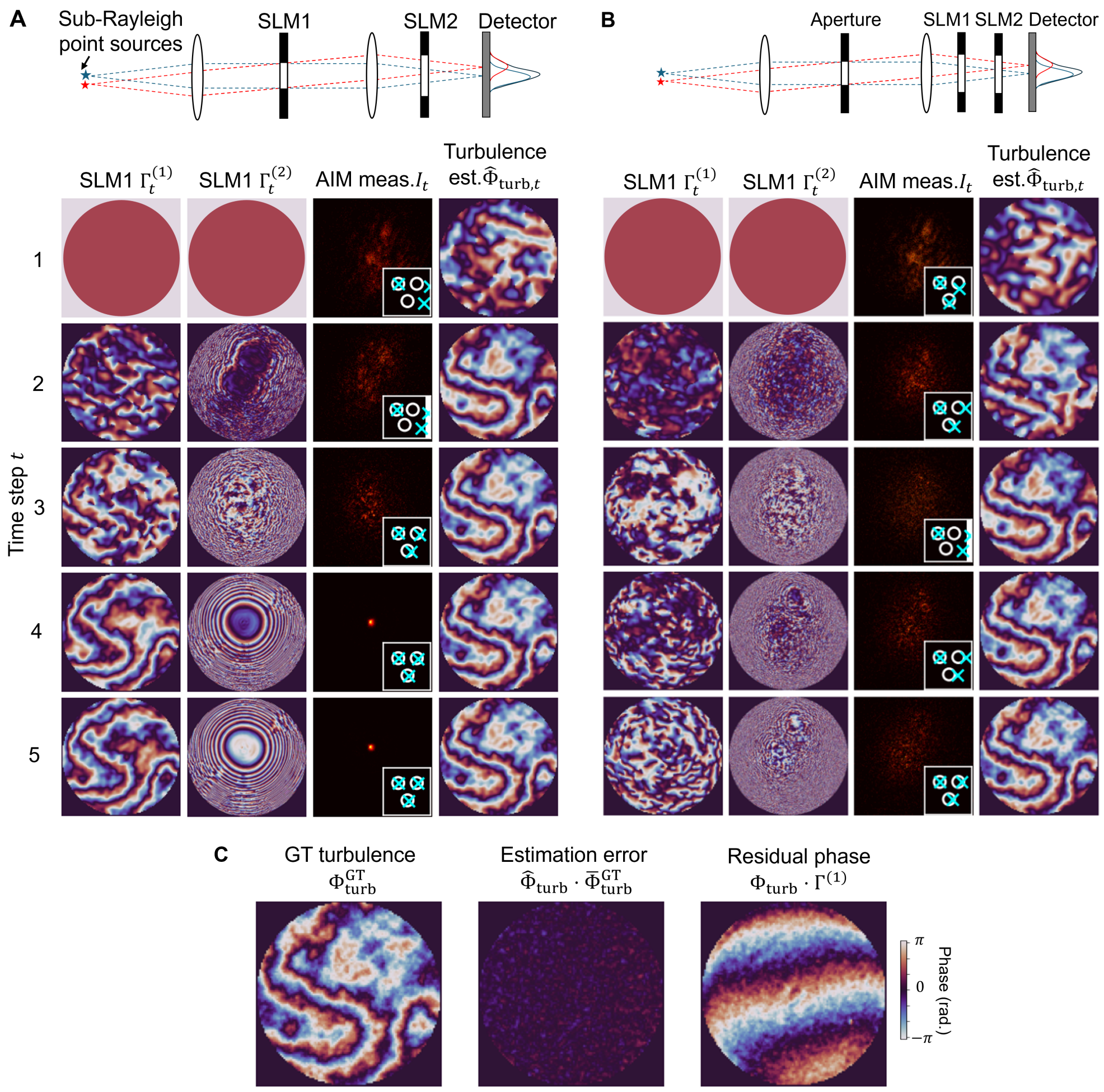


**Fig. S11.** Joint estimation of scene parameters and turbulence under two optical configurations. (A) Configuration 1: SLM1 is positioned at the aperture plane while SLM2 remains downstream of the lens. (B) Configuration 2: Both SLM1 and SLM2 are placed downstream of the lens. For each configuration, five sequential AIM measurements are shown. At each time step $t$, optimized SLM phase patterns ($\Gamma_t^{(1)}$, $\Gamma_t^{(2)}$), the corresponding AIM measurement $I_t$, and the estimated turbulence phase $\hat{\Phi}_{\text{turb},t}$ are displayed. In both configurations, the algorithm successfully recovers the scene parameters together with the turbulence phase. (C) Detailed turbulence analysis for Configuration 1. Left: ground-truth turbulence phase, middle: estimation error, right: residual phase between SLM1 and the ground-truth turbulence. The residual shows that SLM1 effectively compensates the turbulence, leaving primarily a global tip/tilt component. All results are obtained for a three point-source scene at 0.66R separation, with brightness ratios [1, 3, 6] and a total photon budget of 50,000, under turbulence strength with $D/r_0 = 13$.

tions at the pupil. Once the wavefront is stabilized, the second SLM generates the informative, shift-variant PSFs required to resolve closely spaced sources.

In contrast, Configuration 2 does not exhibit this clear decoupling between wavefront correction and information encoding. Because both SLMs are at intermediate planes, pupil-plane turbulence manifests as a complex interplay of amplitude and phase fluctuations at both SLM surfaces. Since the SLMs are restricted to phase-only modulation, the resulting optimized patterns become more intricate, and the measurements appear more scattered compared to Configuration 1. From an experimental standpoint, this complexity implies increased sensitivity to alignment errors and model mismatches.

Consequently, for applications requiring simultaneous turbulence estimation, Configuration 1 offers superior reliability. By placing at least one SLM at the pupil plane, the system naturally exploits the physical hierarchy of the imaging process: correcting global isoplanatic turbulence first (similar to conventional adaptive optics) to provide a stable foundation for the anisoplanatic modulations necessary for super-resolution.

## REFERENCES


1. S. Choi, P. Menart, A. Schramka, *et al.*, “Poisson wavefront imaging in photon-starved scenarios,” J. Opt. Soc. Am. B **43**, 231–238 (2026).
2. S. Choi, P. Menart, A. Schramka, *et al.*, “Photon-starved imaging through turbulence at the diffraction limit,” arXiv preprint arXiv:2510.22806 (2025).
3. H. Choi, S. Choi, P. Menart, *et al.*, “Telescope imaging beyond the rayleigh limit in extremely low snr,” New J. Phys. **26**, 093019 (2024).



\end{document}